\documentclass[reprint,superscriptaddress,amsmath,amssymb,aps,prx]{revtex4-1}
\pdfoutput=1
\usepackage{amsmath}
\usepackage{amsfonts}
\usepackage{amssymb}
\usepackage{amsthm}
\usepackage{bm}
\usepackage{color}
\usepackage{dcolumn}
\usepackage{graphicx}
\usepackage{mathtools}
\usepackage{textgreek}
\usepackage[utf8]{inputenc}

\begin{document}
\title{A Two-Step Biopolymer Nucleation Model Shows a Nonequilibrium Critical Point}

\author{Alexander I. P. Taylor}
\email{ataylor8@sheffield.ac.uk}
\affiliation{Department of Molecular Biology and Biotechnology, University of Sheffield, Sheffield S10 2TN, UK.}

\author{Lianne D. Gahan}
\affiliation{Department of Molecular Biology and Biotechnology, University of Sheffield, Sheffield S10 2TN, UK.}
\affiliation{Department of Physics and Astronomy, University of Sheffield, Sheffield S3 7RH, UK.}

\author{Rosemary A. Staniforth}
\email{r.a.staniforth@sheffield.ac.uk}
\affiliation{Department of Molecular Biology and Biotechnology, University of Sheffield, Sheffield S10 2TN, UK.}

\author{Buddhapriya Chakrabarti}
\email{b.chakrabarti@sheffield.ac.uk}
\affiliation{Department of Physics and Astronomy, University of Sheffield, Sheffield S3 7RH, UK.}

\date{\today}

\setlength{\belowcaptionskip}{-10pt}

\begin{abstract}
Biopolymer self-assembly pathways are central to biological activity, but are complicated by the ability of the monomeric subunits of biopolymers to adopt different conformational states. As a result, biopolymer nucleation often involves a two-step mechanism where the monomers first condense to form a metastable intermediate, and this then converts to a stable polymer by conformational rearrangement of its constituent monomers. While existing mathematical models neglect the dynamics by which intermediates convert to stable polymers, experiments and simulations show that these dynamics frequently occur on comparable timescales to condensation of intermediates and growth of mature polymers, and thus cannot be ignored. Moreover, nucleation intermediates are responsible for cell toxicity in numerous diseases, such as Alzheimer's, Parkinson's, and prion diseases. Due to the relationship between conformation and biological function, the slow conversion dynamics of these species will strongly affect their toxicity. In this study, we present a modified Oosawa model which explicitly accounts for simultaneous assembly and conversion. To describe the conversion dynamics, we propose an experimentally motivated initiation-propagation (IP) mechanism in which the stable phase arises locally within the intermediate, and then spreads through additional conversion events induced by nearest-neighbor interactions, in a manner analogous to one-dimensional Glauber dynamics. Our mathematical analysis shows that the competing timescales of assembly and conversion result in a nonequilibrium critical point, separating a regime where intermediates are kinetically unstable from one where conformationally mixed intermediates can accumulate. In turn, this strongly affects the rate at which the stable biopolymer phase accumulates. In contrast to existing models, this model is able to reproduce commonly observed experimental phenomena such as the formation of mixed intermediates, and abrupt changes in the scaling exponent $\gamma$, which relates the total monomer concentration to the accumulation rate of the stable phase. Our work provides the first step towards a general model of two-step biopolymer nucleation, which can be used to quantitatively predict the concentration and composition of biologically crucial intermediates.
\end{abstract}


\maketitle

\section{Introduction}

Biopolymer formation is essential to life. Uncontrolled self-association of biological molecules to form non-covalent polymers is implicated in diseases such as Alzheimer's \cite{Hardy1992}, Parkinson's \cite{Spillantini1998}, and sickle-cell anemia \cite{Pauling1949}. In many cases, the monomeric subunits of biopolymers adopt distinct conformational states. As a result, nucleation often proceeds via a two-step mechanism, in which the monomers first condense into a metastable intermediate, and the stable polymer develops by conformational conversion of the monomers within this phase \cite{Harper1997,Serio2000,Galkin2007,Chimon2007,Auer2007,Lee2011,Levin2014}. Intermediates have diverse morphologies (Fig.~\ref{fig01}(a)), depending on the monomer interactions which drive condensation. Polymeric intermediates are commonly observed in experimental and computational studies of amyloid formation \cite{Harper1997,Pellarin2007,Urbanc2010,Chatani2015}, and convert to stable polymers via reorganization of their secondary or tertiary structure. In addition, diverse biopolymers including actin and prions are known to nucleate in spheroidal intermediates \cite{Serio2000,Zhu2002,Galkin2007,Chimon2007,Levin2014,Luo2014,Shin2017}. Biopolymer nucleation intermediates are implicated in cell death and resulting disease \cite{Chiti2006,Benilova2012}, and an understanding of their formation is crucial for therapeutic development.

In order to accurately predict the rate and mechanism of two-step biopolymer nucleation, theoretical models must realistically represent the process by which intermediates convert to stable polymers. Existing analytical models treat conversion as an instantaneous process \cite{Vitalis2011,Auer2012,Garcia2014,Saric2016,Lee2017}, neglecting the dynamics by which the stable polymeric phase spreads through the intermediate. However, there is no general reason why the ordering transition should be sufficiently rapid, or cooperative, for this to be valid. In contrast, experimental and computational studies often point to progressive ordering on slow timescales \cite{Chimon2007,Auer2008,Li2008,Kumar2009,Luo2014,Chen2015}, indicating that the stable phase arises locally and spreads through the intermediate. This means that condensation and ordering can occur concurrently \cite{Auer2008,Li2008,Kumar2009,Luo2014}. In addition, simulations support an autocatalytic mechanism by which conformationally ordered monomers promote conversion of their neighbors \cite{Ding2005,Lipfert2005,Auer2008}. This raises the possibility that condensation and ordering may be kinetically coupled, with larger intermediates converting at a higher rate \cite{Auer2012,Saric2016}; however, the kinetic prerequisites for such a mechanism have not been rigorously determined. Thus, in order to describe the time-evolution of intermediate populations and elucidate the role of autocatalysis in two-step nucleation, there is an urgent requirement for a model which explicitly considers the conversion of intermediates by an autocatalytic initiation-propagation (IP) mechanism.

In this paper, we address these issues. We develop a minimal model of two-step nucleation (Fig.~\ref{fig01}(b)-(f)) in which polymeric intermediates undergo an autocatalytic conversion process analogous to one-dimensional (1D) Glauber dynamics \cite{Glauber1963}. Through a combination of analytical calculations and stochastic simulations \footnote{See Supplemental Material at [*URL to be inserted] for a more detailed description of our mathematical analysis and simulations.}, we show that this simple addition to Oosawa's classical model of nucleated polymerization \cite{Oosawa1962,Oosawa1975} produces rich, nonlinear dynamics. In autocatalytic systems, a nonequilibrium critical point arises, separating a regime where intermediates do not accumulate from one where the two-step nature of nucleation becomes apparent. In the vicinity of this critical point, nucleation dynamics resemble a second-order phase transition similar to the two-dimensional (2D) Ising model \cite{Peierls1936,Onsager1944}. It is clear from our derivation that this phase transition does not require a particular dimensionality, and we expect our results to be applicable to morphologically diverse intermediates. Thus, we propose that models of explicit conformational conversion have the capacity to reconcile apparently contradictory interpretations of biopolymer nucleation, and resolve important unanswered questions regarding the origin and nature of metastable intermediates.

\section{Outline of the model}

\begin{figure}[t]
\includegraphics[width=8.5cm]{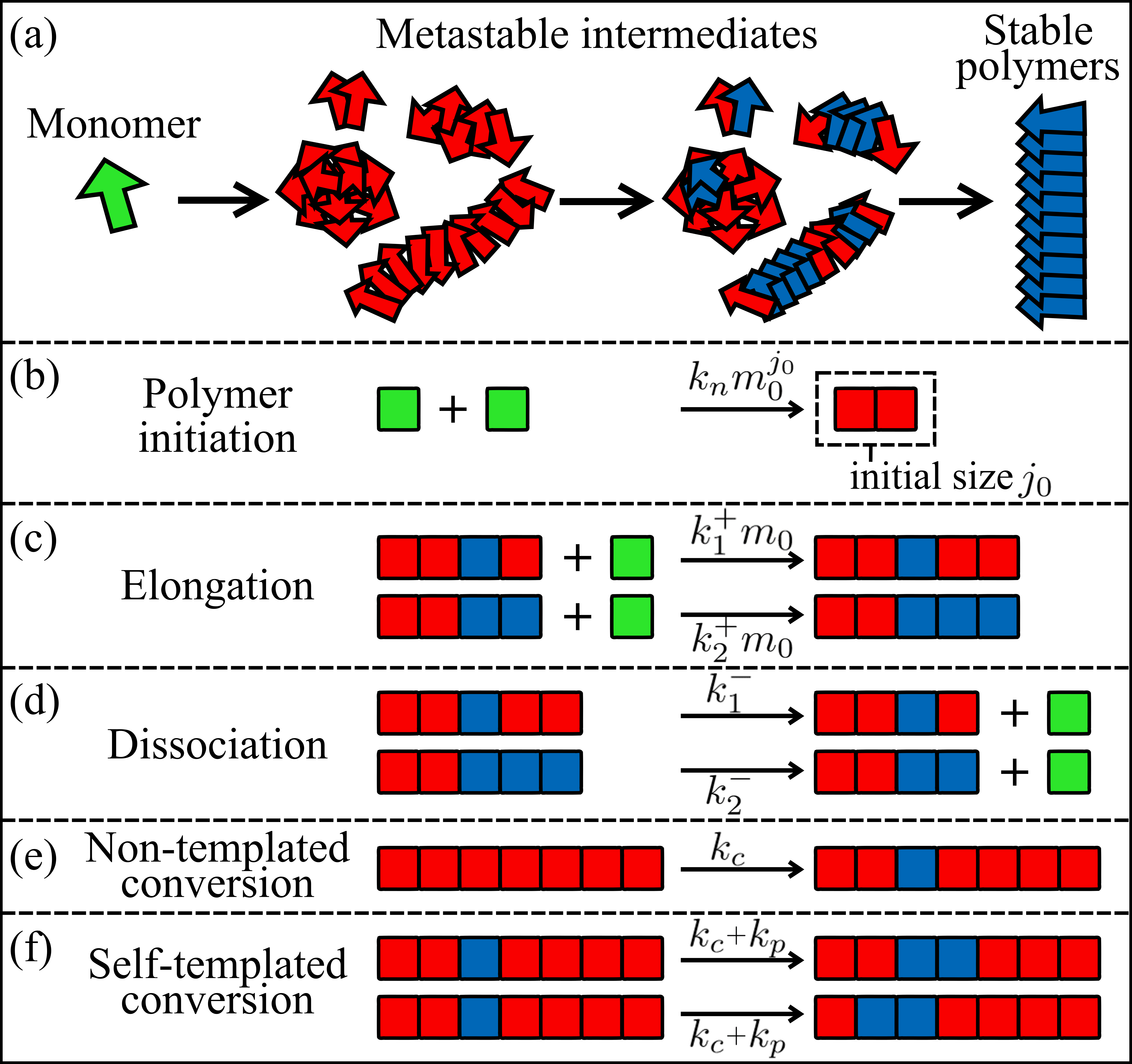} 
\caption{\label{fig01} Schematic of the proposed two-step biopolymer nucleation model. (a) Biopolymer nucleation pathways frequently involve diverse metastable intermediates, which progress to stable polymers by conversion of monomer subunits. (b-f) Schematic representation of our model, accounting for explicit growth and conversion of linear intermediates as described in Sec. II. A. Color scheme for monomers: green, unassembled; red, metastable (type-1); blue, stable (type-2).}
\end{figure}

In this section, we outline the mathematical basis of our two-step biopolymer nucleation model. This model combines the established formalism for nucleated biopolymer self-assembly \cite{Oosawa1962,Oosawa1975} with a proposed initiation-propagation (IP) mechanism of conformational conversion, in which the stable phase arises locally within the intermediate and spreads autocatalytically by nearest-neighbor interactions (Fig.~\ref{fig01}). In this respect, our IP mechanism resembles 1D Glauber dynamics \cite{Glauber1963}. Although the proposed dynamics affect all polymerized monomers, we are particularly interested in their effects on macroscopic, observable nucleated polymerization. Therefore, we concentrate on the aspects of these dynamics which are most relevant to the biopolymer nucleation and growth, and leave a more comprehensive treatment of the internal composition of developing biopolymers as a subject for future work.

\subsection{Microscopic processes}

We consider nucleated polymerization of a free monomer population maintained at a fixed concentration $m_{0}$, far from equilibrium. This situation arises physiologically, when there is a monomer source, and is also valid for experimental systems on timescales shorter than the typical timescale $\tau_{\textrm{inf}}$ for significant monomer depletion to occur, which we discuss in greater detail in Sec. II. B. Each biopolymer is treated as a 1D lattice of monomers with length $j$. A polymerized monomer is in either the metastable (type-1) or stable (type-2) state (Fig.~\ref{fig01}(b)-(f)). Polymerization is initiated at a rate $v_{n}=k_{n}m_{0}^{j_{0}}$, and each polymer initially contains $j_{0}\geq2$ monomers in the type-1 state (Fig.~\ref{fig01}(b)). Polymerized monomers irreversibly convert to the type-2 state at a rate of $k_{c}$, or $k_{c}+k_{p}$ if adjacent to a type-2 monomer (Fig.~\ref{fig01}(e) \& (f)). Thus, $k_{p}$ represents the excess rate due to autocatalysis. For the purpose of this study, it is not necessary to consider cases where a type-1 monomer is bounded by two type-2 monomers, as such cases do not affect the macroscopic kinetics of nucleated polymerization.

Each polymer can elongate by reversible monomer addition at a single `active' end (Fig.~\ref{fig01}(c) \& (d)). Strong bias towards elongation at a single end is often observed experimentally \cite{Goldsbury1999,Ban2004,Sleutel2017}, and is caused by the fact that most biopolymers exhibit structural differences between their ends, which result in contrasting physicochemical properties. Although inclusion of this bias simplifies our analysis, it is not expected to qualitatively alter the results. Therefore, we fully expect our theory to generalize to more unusual cases \cite{Xu2019} where both ends elongate at similar rates. It is also important to note that the forward and reverse elongation rates must exhibit an equal bias, in order to satisfy detailed balance when $m_{0}$ is at the solubility limit. From this point onwards, we simply refer to the forward process as elongation, and the reverse process as dissociation.

During elongation, the conformational state of the monomer being incorporated into the polymer is templated by that of the monomer already at the active end \cite{Serio2000,Straub2011} (Fig.~\ref{fig01}(c) \& (d)). Therefore, if the monomer at the active end is in the type-1 state, then elongation proceeds via addition of type-1 monomers with a forward rate $k_{1}^{+}m_{0}$ and a reverse rate $k_{1}^{-}$. Similarly, if the monomer at the active end is in the type-2 state, then type-2 monomers are added with forward and reverse rates $k_{2}^{+}m_{0}$ and $k_{2}^{-}$, respectively. Thus, polymers exhibit one of two distinct growth modes, type-1 and type-2, corresponding to the conformation of the monomer at the active end and the resulting mechanism of elongation. Growth mode switching occurs when the monomer at the active end converts, causing the emergence of an elongating type-2 phase. Experimental results typically support $k_{2}^{+} \gg k_{1}^{+}$ \cite{Harper1997,Iljina2016}, meaning that the switching rate determines the lifetime of intermediates with a high type-1 structure content, as well as the observed macroscopic kinetics of nucleated polymerization.

\begin{figure}[t]
\includegraphics[width=8.6cm]{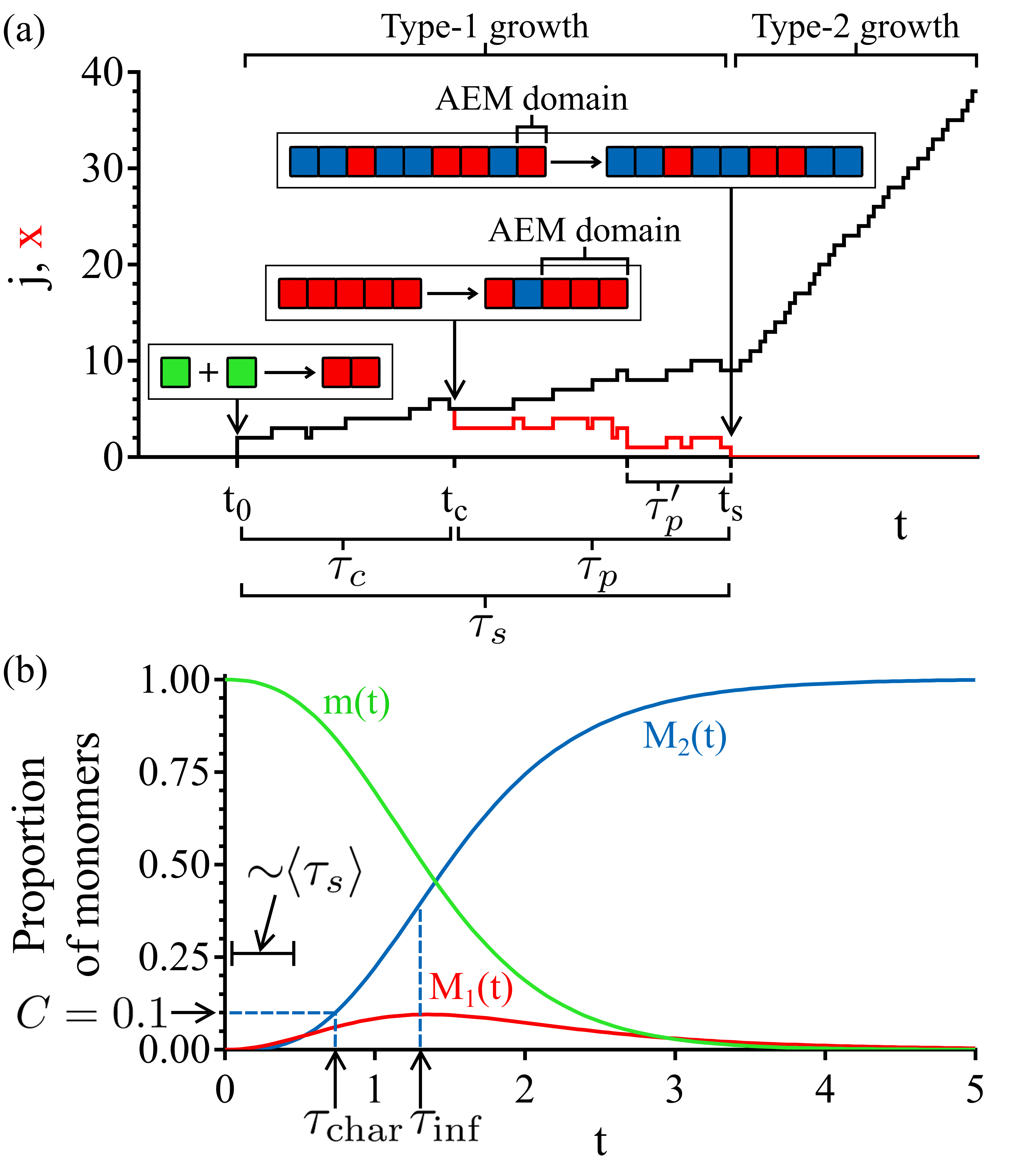} 
\caption{\label{fig02} Characteristic timescales of two-step biopolymer nucleation. (a) Schematic showing an example of how the length ($j$; black line) and AEM domain size ($x$; red line) of a single polymer can change over time. The three key events (Sec. II. B.) of initiation ($t=t_{0}$), conversion of the first monomer ($t=t_{c}$), and growth mode switching ($t=t_{s}$) are annotated with diagrams representing the change in conformational state of the polymer, with the following color scheme for constituent monomers: green, unassembled; red, metastable (type-1); blue, stable (type-2). (b) Schematic showing the macroscopic self-assembly kinetics of a typical polymer population (Sec. II. B.), and relevant characteristic times. Curves represent the proportions of monomers which are unassembled ($m(t)$; green), or incorporated into polymers with a type-1 ($M_{1}(t)$; red) or type-2 ($M_{2}(t)$; blue) growth mode.}
\end{figure}

\subsection{Characteristic timescales}

Before deriving master equations to predict the macroscopic kinetics of nucleated polymerization, it is informative to consider the key conformational transitions that a nascent polymer undergoes, and the timescales on which they occur (Fig.~\ref{fig02}). There are three key events that lead to emergence of a type-2 growth mode: (i) initiation of the polymer at a time $t_{0}$, with all constituent monomers initially in the type-1 state; (ii) non-autocatalytic conversion of the first polymerized monomer at time $t_{c}>t_{0}$, causing the formerly `pure' intermediate to become a mixed intermediate; and (iii) growth mode switching at time $t_{s} \geq t_{c}$, which coincides with conversion of the monomer at the active end (Fig.~\ref{fig02}(a)). If the first monomer to convert is situated at the active end, then $t_{s}=t_{c}$; otherwise, additional conversion events are required for growth mode switching to occur, so that $t_{s}>t_{c}$. If $k_{p} \ll k_{c}$, these additional conversion events will be primarily non-autocatalytic and uncorrelated with one another; if $k_{p} \gg k_{c}$, conversion of the first monomer will stimulate autocatalytic conversion of its neighbours, causing the type-2 state to propagate from its point of origin to the active end. We also identify three important time intervals: (i) the lifetime of a pure intermediate, $\tau_{c}=t_{c}-t_{0}$; (ii) the propagation time $\tau_{p}=t_{s}-t_{c}$, which is the time for a mixed intermediate to adopt a type-2 growth mode; and (iii) the total switching time $\tau_{s}=t_{s}-t_{0}=\tau_{c}+\tau_{p}$, which is the total time taken for an intermediate to acquire a type-2 growth mode (Fig.~\ref{fig02}(a)). In addition, we introduce a single domain wall propagation time $\tau'_{p}$, which is distinct from $\tau_{p}$ and will be discussed in greater detail in Sec. II. C.

We also consider two timescales related to the overall self-assembly of populations of mature polymers: (i) the characteristic self-assembly time $\tau_{\textrm{char}}$, and (ii) the inflection time $\tau_{\textrm{inf}}$ (Fig.~\ref{fig02}(b)). Both are common experimental observables which are derived from phenomenological analysis of biopolymer self-assembly curves, and can be used to derive mechanistic conclusions \cite{Ferrone1999,Cohen2011a}. Specifically, the time-dependent mass of mature polymers follows a convex profile under circumstances where $m_{0}$ is approximately constant. If monomer depletion can occur, the self-assembly curve is sigmoidal and approaches a constant limit due to exhaustion of the free monomer \cite{Ferrone1999} (Fig.~\ref{fig02}(b)). For the purpose of this study, we identify mature polymers as species with a type-2 growth mode, and denote the effective concentration of monomer incorporated into such species as $M_{2}(t)$. The self-assembly time $\tau_{\textrm{char}}$ is the time required for $M_{2}(t)$ to reach an arbitrary, often small proportion of the initial free monomer concentration $m_{0}$, such that $M_{2}(\tau_{\textrm{char}})=Cm_{0}$ and $C$ is a constant in the interval $(0,1)$ \cite{Ferrone1999} (Fig.~\ref{fig02}(b)). While a number of different values of $C$ are used, for small $C$ these different characteristic times typically produce similar scaling behaviors. The inflection time $\tau_{\textrm{inf}}$ is the time at which $M_{2}(t)$ has its inflection point, such that $\partial_{t}^{2}M_{2}(\tau_{\textrm{inf}})=0$ \cite{Cohen2011a} (Fig.~\ref{fig02}(b)). Because inflection is a consequence of monomer depletion, $\tau_{\textrm{inf}}$ provides a general timescale on which monomer depletion occurs. Therefore, the assumption that $m_{0}$ is approximately constant implies that $\tau_{s} \ll \tau_{\textrm{inf}}$, meaning that $M_{2}(t)$ is locally convex.

\subsection{Growth mode switching as an absorbing boundary problem}

The ability of type-2 monomers to autocatalytically induce conversion of their neighbors allows the type-2 state to propagate to the active end from a distant site of origin. Thus, the rate of growth mode switching depends not only on the conformational state of the monomer at the active end, but also the length distribution of the domain of type-1 monomers adjacent to the active end, which we term the `active end metastable' (AEM) domain. When the size $x$ of this domain becomes zero, growth mode switching occurs (Fig.~\ref{fig02}(a)). We are primarily interested in far-from-equilibrium cases where $m_{0} \gg k_{2}^{-}/k_{2}^{+}$, so that growth mode switching has a negligible reverse process and can be treated as an absorbing boundary problem. Initially, the polymer is formed with all monomers in the type-1 state, so that $x=j$ for $t_{0} \leq t < t_{c}$. Following the first conversion event at $t=t_{c}$, the polymer acquires a mixed composition, so that $x<j$ for $t \geq t_{c}$. For $t > t_{c}$, the AEM domain can grow due to elongation at the active end (Fig.~\ref{fig01}(c)), or contract due to either dissociation of monomers from the active end (Fig.~\ref{fig01}(d)) or conversion of monomers within the AEM domain (Fig.~\ref{fig01}(e) \& (f)). Growth mode switching occurs when the AEM domain size hits an absorbing boundary at $x=0$, at the time $t_{s}$.

The fact that both autocatalytic and non-autocatalytic conversion events may occur during propagation leads to a distinction between the total propagation time $\tau_{p}$, and the single domain wall propagation time $\tau'_{p}$ (Fig.~\ref{fig02}(a)). While autocatalytic conversion is restricted to the nearest neighbors of type-2 monomers, non-autocatalytic conversion is not. Thus, non-autocatalytic conversion of type-1 monomers within the AEM domain may cause a new domain wall to form, superseding the existing AEM domain wall. While $\tau_{p}$ denotes the total time between the onset of propagation and growth mode switching, $\tau'_{p}$ denotes the time taken for a single domain wall to successfully propagate to the active end, without being superseded. Therefore, $\tau'_{p}$ only applies to propagation of the last AEM domain wall to form before growth mode switching. If the original domain wall successfully propagates to the active end, then $\tau'_{p}=\tau_{p}$; otherwise, $\tau'_{p} < \tau_{p}$.

\subsection{General master equations}

We consider the multivariate probability mass function (PMF) $p_{j,x}(t; j_{0}, t_{0})$, which gives the probability that a polymer will have total length $j$, AEM domain size $x$, and a type-1 growth mode at time $t$. The master equation for $p_{j,x}(t)$ has the following form:
\begin{equation}
    \begin{split}
        \frac{dp_{j,x}(t)}{dt} =&\;\delta_{j, j_{0}}\delta_{x, j_{0}}\delta(t-t_{0})\\
        +&\;k_{c}\sum_{y=x+1}^{j}p_{j,y}(t) - k_{c}xp_{j,x}(t)\\
        +&\;k_{p}[\theta(j\text{-}x\text{-}2)p_{j,x+1}(t)-\theta(j\text{-}x\text{-}1)p_{j,x}(t)]\\
        +&\;k_{1}^{+}m_{0}[p_{j-1,x-1}(t)-p_{j,x}(t)]\\
        +&\;k_{1}^{-}[p_{j+1,x+1}(t)-p_{j,x}(t)],
    \end{split}
\end{equation}
where $\delta_{j, j_{0}}$ and $\delta(t-t_{0})$ are the Kronecker and Dirac deltas, respectively, and $\theta(z)$ is a step function such that $\theta(z)=1$ if $z \geq 0$ and $\theta(z)=0$ if $z<0$. As with previous models \cite{Knowles2009,Cohen2011a,Michaels2014}, $j_{0}$ is taken to be a minimum polymer length below which rapid disassembly almost always occurs; therefore, an absorbing boundary exists such that $p_{j_{0}-1,x}(t)=0$. In addition, to account for irreversible growth mode switching, we impose the absorbing boundary condition $p_{j,0}(t)=0$. While the first term in Eq. (1) accounts for initiation at $t=t_{0}$, the terms proportional to $k_{c}$, $k_{p}$, $k_{1}^{+}m_{0}$, and $k_{1}^{-}$ account for non-autocatalytic conversion, autocatalytic conversion, elongation, and dissociation, respectively.

The terms accounting for non-autocatalytic conversion introduce a jump process to Eq. (1), which complicates efforts to satisfy the boundary conditions. Nonetheless, analytical solutions can be obtained in specific cases. In the case where $x=j$, corresponding to the length distribution of intermediates with a purely type-1 composition, the summation disappears and the problem becomes significantly more tractable. We use this special case to obtain $\langle\tau_{c}\rangle$ in Sec. III. B. For mixed intermediates, the time-dependence of AEM domain size can be solved by summing over the polymer length. Let $\phi_{x}(t; x_{c}, t_{c})$ represent the probability that a mixed intermediate has an AEM domain of size $x$ and a type-1 growth mode at time $t$, given the boundary condition that $x=x_{c}$ when $t=t_{c}$. We can determine $\phi_{x}(t)$ by summing $p_{j,x}(t)$ over $j$:
\begin{equation}
    \phi_{x}(t) = \sum_{j=j_{0} \lor x+1}^{\infty}p_{j,x}(t),
\end{equation}
where $j_{0} \lor x+1$ denotes the maximum of $j_{0}$ and $x+1$. The corresponding master equation is as follows:
\begin{equation}
\begin{split}
    \frac{d\phi_{x}(t)}{dt} =&\;\delta_{x,x_{c}}\delta(t-t_{c}) + k_{c}\sum_{y=x+1}^{\infty}\phi_{y}(t)-k_{c}x\phi_{x}(t)\\
    +&\;k_{1}^{+}m_{0}[\phi_{x-1}(t)-\phi_{x}(t)]\\
    +&\;(k_{1}^{-}+k_{p})[\phi_{x+1}(t)-\phi_{x}(t)].
\end{split}
\end{equation}
As with related theories \cite{Ferrone1999,Knowles2009,Cohen2011a}, we consider the case where $k_{1}^{-}$ is small, allowing losses due to destabilization of polymers with length $j<j_{0}$ to be neglected. The absorbing boundary at $x=0$ corresponds to growth mode switching, and is retained. Eq. (3) is the discrete analog of a jump-diffusion equation:
\begin{equation}
\begin{split}
    \frac{\partial\phi(x,t)}{\partial t} =&\;\delta(x-x_{c})\delta(t-t_{c}) + k_{c}\int_{x}^{\infty}\phi(y,t)dy\\
    -&\;\Big(k_{c}x + \mu\frac{\partial}{\partial x} - D\frac{\partial^{2}}{\partial x^{2}}\Big)\phi(x,t),
\end{split}
\end{equation}
where $\mu=k_{1}^{+}m_{0}-k_{1}^{-}-k_{p}$ and $D=(k_{1}^{+}m_{0}+k_{1}^{-}+k_{p})/2$ are the drift and diffusion coefficients for AEM domain size, with $D \geq 0$ and $-2D\leq\mu\leq2D$. Because Eq. (3) can be written in an equivalent form to Eq. (4), $\mu$ and $D$ are also informative in a discrete context. We define an effective P\'eclet number $x_{c}|\mu|/D$ for propagation of the type-2 state to the active end. Propagation is primarily diffusive when $x_{c}|\mu|/D < 1$, and advective when $x_{c}|\mu|/D > 1$. In the advective case, the AEM domain tends to contract if $\mu < 0$, and expand if $\mu > 0$. Thus, $\mu$ is the expected rate of AEM domain expansion due to the competing effects of type-1 growth and type-2 propagation, and this competition provides the basis of the critical behavior we observe in this paper.

\section{Critical phenomena in growth mode switching}

In this section, we investigate how the competing effects of propagation and elongation on AEM domain size alter the lifetime of intermediates. Using a combination of stochastic simulations and analytical theory, we uncover the existence of a nonequilibrium critical point resulting from transcritical bifurcation of an order parameter $P_{e}$ around the point $\mu=0$. This critical point shares many common features with equilibrium second-order phase transitions, including a jump in $\partial_{\mu}P_{e}$, critical slowing, and diverging susceptibility to autocatalytic effects. These effects will be shown in the following section to significantly affect the macroscopic kinetics of two-step biopolymer self-assembly.

\begin{figure*}[t]
\includegraphics[width=17.7cm]{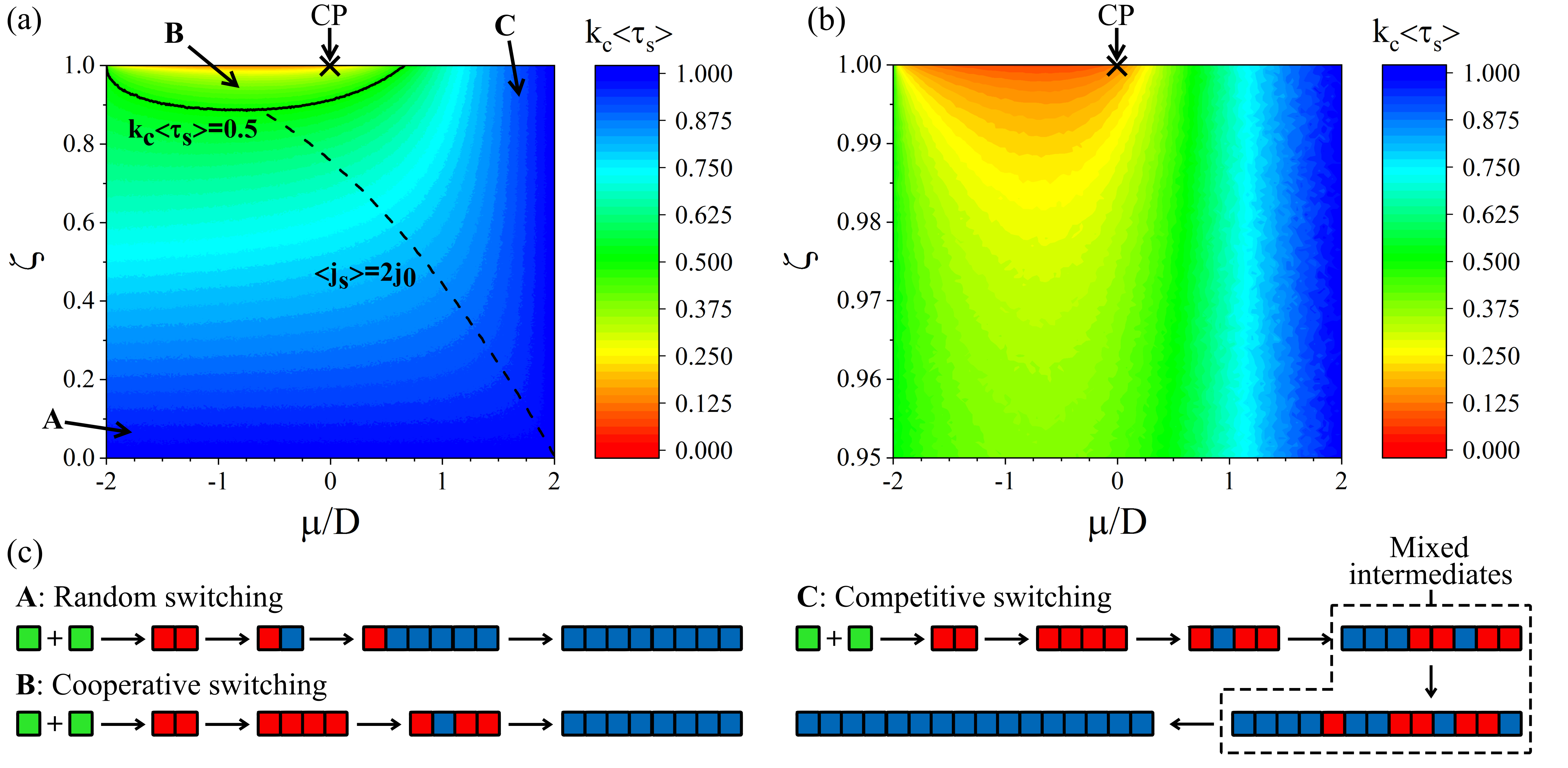} 
\caption{\label{fig03} Two-step biopolymer nucleation exhibits distinct dynamical regimes. (a) Nonequilibrium phase diagram superimposed on a heat map of mean growth mode switching time $k_{c}\langle\tau_{s}\rangle$ (color scale). The solid and dashed lines are contours which separate three dynamical regimes (A, random switching; B, cooperative; C, competitive). At the critical point (CP), $\partial\langle\tau_{s}\rangle/\partial(\mu/D)$ undergoes a jump. (b) A close-up of the heat map of $k_{c}\langle\tau_{s}\rangle$ at high $\zeta$ values ($0.95\leq\zeta<1.0$). Varying $\mu/D$ or $\zeta$ close to the critical point causes a rapid change in the switching time. (c) Schematics of regimes A-C, with color representing monomer conformation: green, unassembled; red, metastable (type-1); blue, stable (type-2).}
\end{figure*}

\subsection{Existence of a nonequilibrium critical point}

To evaluate how the competing effects of propagation and elongation on AEM domain size alter the lifetime of intermediates, we calculate the mean switching time $\langle\tau_{s}\rangle$. For this purpose, we use a rejection-free kinetic Monte Carlo (rfKMC; see Supplemental Material at \footnotemark[\value{footnote}]) algorithm to determine hitting times at the boundary $x=0$ in Eq. (1). The rfKMC algorithm converges to the exact solution for the dynamics as the number of polymers $N\to\infty$ \cite{Bortz1975,Gillespie1976,Voter2005}, and its application to biochemical problems including biopolymer self-assembly is well-established \cite{Gillespie1976,Morris2013,Eden2015}. In this case, the rfKMC algorithm is preferable to deterministic methods as it samples broad polymer size distributions more efficiently. For the purpose of the simulations, we set $k_{1}^{-}=0$ so that there is a single absorbing boundary, although our subsequent mathematical analysis (Sec. III. C. \& D.; \footnotemark[\value{footnote}]) is more generally applicable to small, non-zero $k_{1}^{-}$. In addition, while we have carried out simulations for diverse $j_{0}$, the results are qualitatively similar in all cases \footnotemark[\value{footnote}]; therefore, we focus on the $j_{0}=2$ case as a representative example. To explore the system's limiting behavior in a 2D space, it is convenient to normalize both the timescales and rate constants with respect to $k_{c}$. Therefore, we use relative time $k_{c}t$ and sample points in $(k_{p}/k_{c},\,k_{1}^{+}m_{0}/k_{c})$ space by varying a pair of bounded, mechanistically significant parameters, $\mu/D$ and $\zeta=k_{p}/(k_{c}+k_{p})$. While $\mu/D$ defines the tendency for expansion or contraction of the AEM domain (Sec. II. D.), $\zeta$ is the proportion of the conversion rate adjacent to a domain wall which is attributable to autocatalytic effects. Thus, $\zeta$ provides a measure of the importance of autocatalysis in conformational conversion of the assembled monomers.

In Fig.~\ref{fig03}(a), we show how the normalized mean growth mode switching time $k_{c}\langle\tau_{s}\rangle$ varies as a function of the model parameters $(\mu/D, \zeta)$. We use these data to plot a nonequilibrium phase diagram, which we divide into three different dynamical regimes (\footnotemark[\value{footnote}]; Fig.~\ref{fig03}(c)): (A) random, (B) cooperative, and (C) competitive. In the random switching regime, elongation and autocatalysis do not occur, so switching depends entirely on non-autocatalytic conversion of the monomer at the active end. In the cooperative regime ($\zeta\to1$ when $\mu/D\lessapprox0$) (Fig.~\ref{fig03}), autocatalytic conversion events allow the type-2 state to spread from a distant site of origin to the active end. Thus, non-autocatalytic conversion events occurring anywhere in the intermediate can indirectly lead to growth mode switching, causing the instantaneous switching rate to approach $k_{c}j$. This means that mass accumulation and autocatalytic conversion cooperate to bring about a change in growth mode. In the competitive regime, which arises for $\mu/D \gtrapprox 0$, the effects of elongation on AEM domain size outweigh those of autocatalytic conversion. Therefore, the AEM domain undergoes net expansion, which inhibits propagation of the type-2 state to the active end. This causes an increase in $k_{c}\langle\tau_{s}\rangle$, allowing accumulation of large, metastable intermediates with a mixed composition. In this regime, mass accumulation and autocatalytic conversion can be said to compete with one another. Although regimes A-C are continuous with one another, we observe a singular point at $(\mu/D, \zeta)=(0,1)$, where the second derivative of $\langle\tau_{s}\rangle$ with respect to $\mu/D$ diverges in a manner resembling a second-order phase transition (Fig.~\ref{fig03}(a) \& (b),~\ref{fig05}(a)).

\subsection{Lifetime of pure intermediates}

\begin{figure*}[t]
\includegraphics[width=17.7cm]{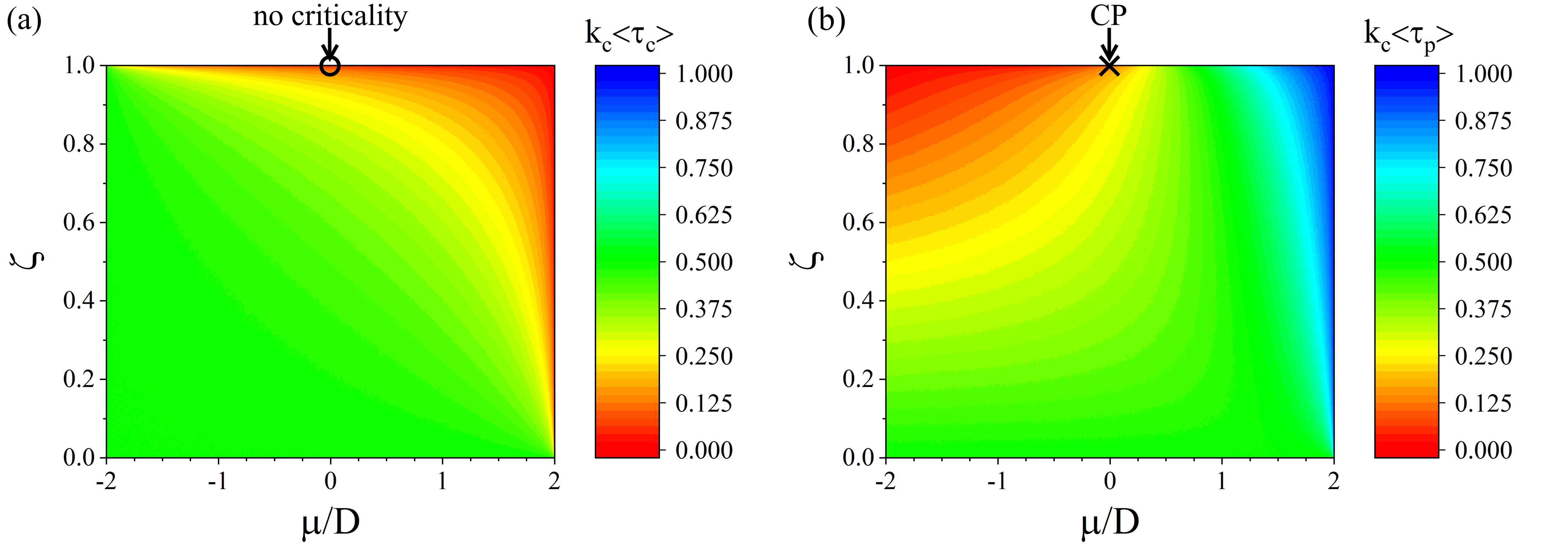} 
\caption{\label{fig04} The critical point in $\langle\tau_{s}\rangle=\langle\tau_{c}\rangle+\langle\tau_{p}\rangle$ is caused by a similar critical point in $\langle\tau_{p}\rangle$, but not $\langle\tau_{c}\rangle$. (a) Heat map of $k_{c}\langle\tau_{c}\rangle$, the normalized mean lifetime of pure intermediates. The phase behavior is described by Eq. (6), and does not exhibit a critical point at $(\mu/D, \zeta)=(0, 1)$. (b) Heat map of $k_{c}\langle\tau_{p}\rangle$, the normalized mean time for mixed intermediates to undergo growth mode switching. Mixed intermediates are kinetically unstable ($k_{c}\langle\tau_{p}\rangle=0$) along the line between $(\mu/D, \zeta)=(-2, 1)$ and $(0, 1)$, and it follows that $\partial\langle\tau_{p}\rangle/\partial(\mu/D)=0$ along this line as well. A second-order critical point (CP) occurs at $(\mu/D, \zeta)=(0, 1)$, where $\partial\langle\tau_{p}\rangle/\partial(\mu/D)$ jumps to a non-zero value.}
\end{figure*}

The switching time $\langle\tau_{s}\rangle$ must reflect a similar critical point in the lifetime of pure intermediates $\langle\tau_{c}\rangle$, or the propagation time of mixed intermediates $\langle\tau_{p}\rangle$, since $\langle\tau_{s}\rangle=\langle\tau_{c}\rangle+\langle\tau_{p}\rangle$. The fact that the critical point occurs on the $\mu=0$ line, along which the competing effects of autocatalytic conversion and elongation on AEM domain size are balanced, indicates that an effect on propagation is responsible for the observed criticality. To ascertain whether $\langle\tau_{c}\rangle$ also plays a role in the critical point, we solve Eq. (1) in the case $x=j$ to obtain a moment-generating function (MGF) for $\tau_{c}$. As our simulations were carried out in the $k_{1}^{-}=0$ case, we apply the same condition when determining $\langle\tau_{c}\rangle$. The MGF of the first-conversion time is given by $\langle e^{-s\tau_{c}} \rangle = 1 - s\sum\hat{p}_{j,j}(s)$, where $\hat{p}_{j,j}(s)=\int_{t_{0}}^{\infty}p_{j,j}(t)e^{-s(t-t_{0})}dt$ is the Laplace transform of $p_{j,j}(t)$. By taking the Laplace transform of Eq. (1) in the case $x=j$ and rearranging, we obtain the following result \footnotemark[\value{footnote}]:
\begin{equation}
\begin{split}
    \langle e^{-s\tau_{c}} \rangle
    =&\;1-\frac{s}{k_{c}}e^{\frac{k_{1}^{+}m_{0}}{k_{c}}}\Big(\frac{k_{1}^{+}m_{0}}{k_{c}}\Big)^{-\big(\frac{s+k_{1}^{+}m_{0}}{k_{c}}+j_{0}\big)}\\
    \times&\;\gamma\Big(\frac{s+k_{1}^{+}m_{0}}{k_{c}}+j_{0}, \frac{k_{1}^{+}m_{0}}{k_{c}}\Big),
\end{split}
\end{equation}
where $\gamma$ is a lower incomplete gamma function. Thus,
\begin{equation}
\begin{split}
    \langle\tau_{c}\rangle
    =&\;\frac{1}{k_{c}}e^{\frac{k_{1}^{+}m_{0}}{k_{c}}}\Big(\frac{k_{1}^{+}m_{0}}{k_{c}}\Big)^{-\big(\frac{k_{1}^{+}m_{0}}{k_{c}}+j_{0}\big)}\\
    \times&\;\gamma\Big(\frac{k_{1}^{+}m_{0}}{k_{c}}+j_{0}, \frac{k_{1}^{+}m_{0}}{k_{c}}\Big).
\end{split}
\end{equation}
Eq. (6) decreases monotonically with $k_{1}^{+}m_{0}/k_{c}$, and has the limits $\langle\tau_{c}\rangle\to(k_{c}j_{0})^{-1}$ as $k_{1}^{+}m_{0}/k_{c}\to0$, and $\langle\tau_{c}\rangle\to0$ as $k_{1}^{+}m_{0}/k_{c}\to\infty$. This behavior reflects the fact that elongation increases the number of monomers within the polymer which can non-autocatalytically convert, reducing the time until the first monomer converts. Thus, if growth does not occur ($k_{1}^{+}m_{0}/k_{c}\to0$), then the total conversion rate is simply $k_{c}j_{0}$. Conversely, if growth is rapid, then the length goes to infinity and the time until the first conversion event becomes negligible. Importantly, Eq. (6) does not predict any form of criticality, and the fact that it does not depend on $k_{c}$ means that it cannot be expressed as a function of $\mu/D$ and $\zeta$ alone. This prediction is confirmed by analysis of $k_{c}\langle\tau_{c}\rangle$ values from our rfKMC trajectories, which we present in Fig.~\ref{fig04}(a). Given that $k_{1}^{+}m_{0}/k_{c}=[\zeta(2+\xi)]/[(1-\zeta)(2-\xi)]$, the phase behavior of $k_{c}\langle\tau_{c}\rangle$ in Fig.~\ref{fig04}(a) is accounted for by Eq. (6) alone, with no critical point. Thus, changes in $\langle\tau_{p}\rangle$ rather than $\langle\tau_{c}\rangle$ must be responsible for the observed criticality.

\subsection{Nature of the critical point}

\begin{figure*}[t]
\includegraphics[width=17.7cm]{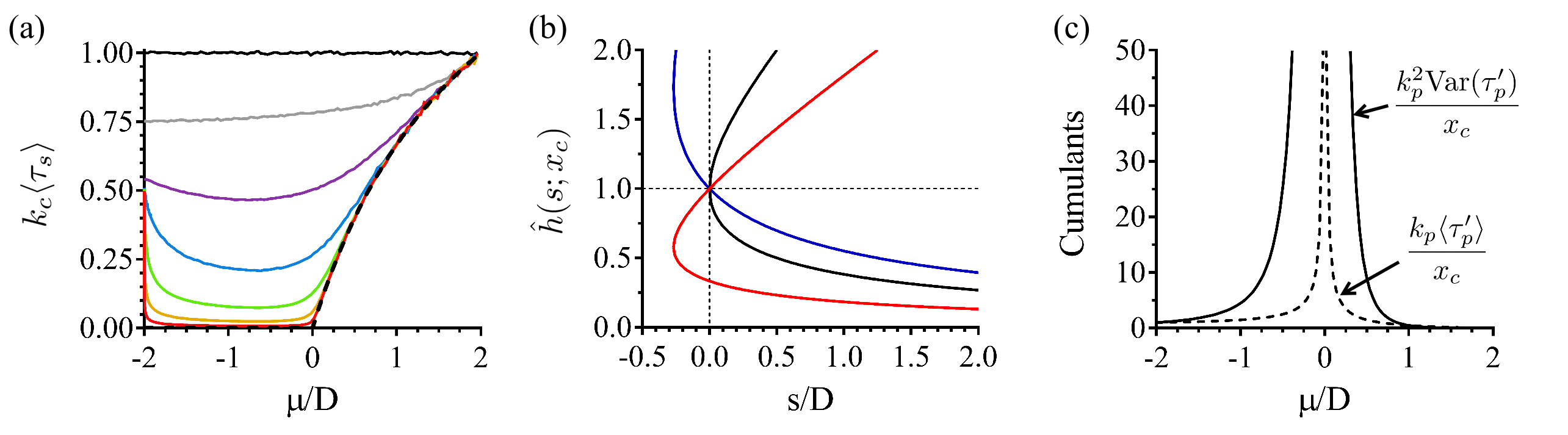} 
\caption{\label{fig05} A continuous phase transition arises in the limit $\zeta=k_{p}/(k_{c}+k_{p}) \to 1$. (a) Dependence of $k_{c}\langle\tau_{s}\rangle$ on $\mu/D$ for varying $k_{p}$. Color scale: solid black, $k_{p}=0$; gray, $k_{p}=k_{c}$; purple, $k_{p}=10k_{c}$; blue, $k_{p}=10^{2}k_{c}$; green, $k_{p}=10^{3}k_{c}$; amber, $k_{p}=10^{4}k_{c}$; red $k_{p}=10^{5}k_{c}$; dashed black, asymptotic scaling law from Eq. (9), showing emergence of a critical point at $\mu/D=0$. (b) Origin of the criticality, showing an exchange of intercepts by the physical and non-physical branches of $\hat{h}(s; x_{c})$ in the $\zeta\to1$ limit. Color scale: blue, $\mu=-D$; black, $\mu=0$; red, $\mu=D$. (c) Divergence of the mean ($k_{p}\langle\tau'_{p}\rangle/x_{c}$; dashed) and variance ($k_{p}^{2}\textrm{Var}(\tau'_{p})/x_{c}$; solid) of the single domain wall propagation time around the critical point in the $\zeta \to 1$ limit, normalized with respect to $k_{p}$ and $x_{c}$.}
\end{figure*}

The critical point in $\langle\tau_{s}\rangle$ must be caused by an abrupt change in the manner in which the type-2 state propagates to the active end. This deduction is supported by analysis of $k_{c}\langle\tau_{p}\rangle$ values from our rfKMC trajectories, which we present in Fig.~\ref{fig04}(b). While $k_{c}\langle\tau_{c}\rangle$ is unaffected by the critical point, $k_{c}\langle\tau_{p}\rangle$ exhibits second-order critical behavior similar to $k_{c}\langle\tau_{s}\rangle$. Specifically, mixed intermediates are kinetically unstable ($k_{c}\langle\tau_{p}\rangle=0$) along a line from $(\mu/D, \zeta)=(-2, 1)$ to $(0, 1)$, corresponding to the idealized case in the autocatalytic limit where propagation of the AEM domain wall to the active end is instantaneous. Because the value of $k_{c}\langle\tau_{p}\rangle$ is constant along this line, it follows that $\partial\langle\tau_{p}\rangle/\partial(\mu/D)=0$ along this line as well. At the critical point at $(\mu/D, \zeta)=(0, 1)$, we observe a jump in $\partial\langle\tau_{p}\rangle/\partial(\mu/D)=0$, allowing $k_{c}\langle\tau_{p}\rangle$ to take nonzero values to the right of this point. To understand this result, let us consider the propagation of a single AEM domain wall from its site of origin to the active end. For this purpose, we solve Eq. (3) in the case where $k_{c}x \ll k_{p}$, so that the terms proportional to $k_{c}$ become negligible. This describes the propagation of a single domain wall when the probability of additional non-autocatalytic conversion events is negligible ($\zeta\to1$), and $x$ remains finite. We obtain $\phi_{x}(t; x_{c}, t_{c})$ using lattice Fourier transforms and the image method (see Supplemental Material at \footnotemark[\value{footnote}]):
\begin{equation}
\begin{split}
    \phi_{x}(t) =&\;e^{-2D(t-t_{c})}\Big(\frac{2D+\mu}{2D-\mu}\Big)^{\frac{x-x_{c}}{2}}\\
    \times&\;\big\{I_{|x-x_{c}|}[\nu(t-t_{c})]-I_{|x+x_{c}|}[\nu(t-t_{c})]\big\},
\end{split}
\end{equation}
for $x \geq 0$ and $t \geq t_{c}$, where $I_{x}(t)$ is a modified Bessel function of the first kind and $\nu=\sqrt{4D^{2}-\mu^{2}}$. The switching rate, which is equal to the hitting rate at the boundary $x=0$, is given by $h(t; x_{c}, t_{c})=(k_{1}^{-}+k_{p})\phi_{1}(t; x_{c}, t_{c})$. We define an escape probability $P_{e}(x_{c})=\lim_{t\to\infty}\sum_{x}\phi_{x}(t; x_{c}, t_{c})$, which is the probability that a given AEM domain wall does not reach the active end as $t\to\infty$. In the $\zeta\to1$ limit, this can only occur if $x$ goes to infinity. In this case, additional non-autocatalytic conversion events become non-negligible, allowing propagation to be interrupted by formation of a new domain wall closer to the active end. Thus, $P_{e}(x_{c})$ is the probability that the AEM domain does not disappear without additional non-autocatalytic conversion events occurring. From final value theorem, we find that $P_{e}(x_{c})=1-\lim_{s\to0^{+}}\hat{h}(s; x_{c})$. The Laplace transform $\hat{h}(s; x_{c})=\int_{t_{c}}^{\infty}h(t; x_{c}, t_{c})\exp[-s(t-t_{c})]dt$ has the following form \footnotemark[\value{footnote}]:
\begin{equation}
    \hat{h}(s; x_{c}) = \Big(\frac{s+2D-r_{\pm}}{2D+\mu}\Big)^{x_{c}},
\end{equation}
where $r_{\pm}=\pm\sqrt{s^{2}+4Ds+\mu^{2}}$. As shown in Fig.~\ref{fig05}(b), Eq. (8) delineates a parabola-like curve, which opens to the right and whose upper ($r_{-}$, unphysical) and lower ($r_{+}$, physical) branches meet at the point $(s^{\ast},\hat{h}^{\ast})$, where $s^{\ast}=\nu-2D\leq0$. When $\mu \neq 0$, both branches intercept the $\hat{h}$ axis; when $\mu=0$, $(s^{\ast},\hat{h}^{\ast})$ touches the $\hat{h}$ axis and the branches exchange intercepts, causing an abrupt change in $\partial_{\mu}P_{e}(x_{c})$:
\begin{equation}
    P_{e}(x_{c}) =
        \begin{cases}
        0,&\mu\leq0,\\
        1-\big(\frac{2D-\mu}{2D+\mu}\big)^{x_{c}},&\mu>0.
        \end{cases}
\end{equation}
In the $\zeta\to1$ limit, autocatalytic conversion and growth occur on much faster timescales than non-autocatalytic conversion. Therefore, the mean switching time is asymptotically $k_{c}\langle\tau_{s}\rangle\sim\lim_{j\to\infty}\big\{\sum_{x_{c}=0}^{j-1}[1-P_{e}(x_{c})]\big\}^{-1}=P_{e}(1)$ \footnotemark[\value{footnote}]. Thus, Eq. (9) explains the critical point in $k_{c}\langle\tau_{s}\rangle$, and the agreement between our calculated scaling law and the simulated result is shown in Fig.~\ref{fig05}(a). The emergence of the critical point in the $\zeta\to1$ limit (Fig.~\ref{fig05}(a)) is remarkably similar to the behavior of the 2D Ising model in the case of vanishing magnetic field $H$ \cite{Onsager1944}. Therefore, we propose that the appearance of nonzero $P_{e}(x_{c})$ for $\mu>0$ represents a second-order phase transition, where $P_{e}(x_{c})$ is an order parameter and $(\mu,\zeta)$ play an analogous role to $(T,H)$ in the Ising model. This transition is associated with the emergence of a regime where mixed intermediates have a finite lifetime, and their continued formation allows a nonequilibrium steady-state population to persist.

\subsection{Diverging propagation times and susceptibility at the critical point}

As further evidence for this critical point, we observe that the moments of the propagation time diverge when $(\mu/D, \zeta)=(0,1)$ (Fig.~\ref{fig05}(c)). As discussed in Sec. II. C., $\tau'_{p}$ is the time taken for a single domain wall to successfully propagate to the active end, without being interrupted by formation of a new domain wall closer to the end (Fig.~\ref{fig02}(a)). In the $\zeta\to1$ limit, the $n^{th}$-order moment is given by $\langle(\tau'_{p})^{n}\rangle=(-1)^{n}\lim_{s\to0^{+}}(\partial_{s}^{n}\hat{h})/\hat{h}$. The upper and lower branches of $\hat{h}$ meet at $s^{\ast}=0$ when $\mu=0$, so the derivatives $\partial_{s}^{n}\hat{h}$ diverge and the moments behave correspondingly. More generally,
\begin{subequations}
\begin{eqnarray}
    \lim_{\zeta\to1}\langle\tau'_{p}\rangle&=&\frac{x_{c}}{|\mu|},\\
    \lim_{\zeta\to1}\textrm{Var}(\tau'_{p})&=&\frac{2x_{c}D}{|\mu|^{3}},
\end{eqnarray}
\end{subequations}
which strongly resembles the diverging magnetic susceptibility $\chi_{H}=\partial_{H}M$ near the Curie point. Because $\tau'_{p}$ is the time interval in which non-autocatalytic conversion events can interrupt propagation, a connection to a susceptibility of the form $\chi_{\zeta}\propto\partial_{\zeta}P_{e}$ is highly intuitive. To test this connection, we determine analytic bounds for the susceptibility in the $\zeta\to1$ limit. We define the susceptibility as $\chi_{\zeta}=-\partial_{\zeta}P_{e}=k_{p}\partial_{k_{c}}P_{e}$, where the sign of our definition ensures a positive value, but does not affect the presence or absence of a divergence. This is essentially the effect of $\zeta$ on the probability that propagation will be interrupted by a non-autocatalytic conversion event in the AEM domain. By differentiating Eq. (3) with respect to $k_{c}$, we obtain a master equation which can be used to determine $\partial_{k_{c}}\phi_{1}(t, k_{c}; x_{c}, t_{c})$ \footnotemark[\value{footnote}]. In turn, this can be used to assess the impact of $k_{c}$ on the hitting rate and $P_{e}$. We solve this master equation to obtain upper and lower bounds for $\partial_{k_{c}}\phi_{1}(t, k_{c}; x_{c}, t_{c})$:
\begin{subequations}
\begin{align}
    \chi_{\zeta}\,=&\,\;C(\mu, D, x_{c})\frac{k_{p}(1-P_{e})}{2}\langle\tau'_{p}\rangle,\\
    1\,\leq\,C(\mu&, D, x_{c})\,\leq\,x_{c}\bigg(\frac{2D}{|\mu|}\bigg) + \bigg(\frac{2D}{|\mu|}\bigg)^{2}.
\end{align}
\end{subequations}
Thus, there is a divergence in $\chi_{\zeta}$ at the critical point which has a scaling law between $|\mu|^{-1}$ and $|\mu|^{-3}$ and is closely related to the divergence of $\langle\tau'_{p}\rangle$ and $\textrm{Var}(\tau'_{p})$.

\subsection{Relationship to a transcritical bifurcation}

The exchange of intercepts by the branches of $\hat{h}(s; x_{c})$ closely resembles a transcritical bifurcation (Fig.~\ref{fig05}(b)), raising the question of whether $P_{e}(x_{c})$ can be written as a fixed point of the corresponding normal form. We begin by noting that the escape probability can be expressed as a limit $P_{e}(x_{c})=\lim_{t\to\infty}\Phi(t; x_{c}, t_{c})$ of the probability $\Phi(t; x_{c}, t_{c})=\sum_{x}\phi_{x}(t; x_{c}, t_{c})$ that the AEM domain wall has not yet propagated to the active end, under conditions where $k_{c}$ is sufficiently small that additional non-autocatalytic conversion events are negligible. Because $P_{e}(x_{c})=1-[1-P_{e}(1)]^{x_{c}}$, a bifurcation in $P_{e}(1)$ must cause a similar behavior in $P_{e}(x_{c})$ \footnotemark[\value{footnote}]; therefore, we focus on the case where $x_{c}=1$. Using the conservation law $\partial_{t}\Phi(t; x_{c}, t_{c}) = \delta(t-t_{0}) - h(t; x_{c}, t_{c})$, we re-express Eq. (8) in the form:
\begin{equation}
    s^{2}\hat{\Phi}(s; 1) - s = \mu s\hat{\Phi}(s; 1) - k_{1}^{+}m_{0}s^{2}\hat{\Phi}(s; 1)^{2}.
\end{equation}
Either by applying final value theorem and the power rule of limits \cite{Bartsch1974} to Eq. (12), or by taking the inverse Laplace transform and evaluating the convolution that corresponds to the $\hat{\Phi}(s; 1)^{2}$ term, it is possible to obtain the following result in the $t\to\infty$ limit \footnotemark[\value{footnote}]:
\begin{equation}
    \frac{d\Phi(t; 1, t_{c})}{dt} \approx \mu\Phi(t; 1, t_{c}) - k_{1}^{+}m_{0}\Phi(t; 1, t_{c})^{2}.
\end{equation}
This is the normal form of a transcritical bifurcation. The physical and non-physical branches of Eq. (8) correspond to the regions of Eq. (13) where $\partial_{t}\Phi(t; 1, t_{c})$ decreases or increases with $\Phi(t; 1, t_{c})$, respectively, and the exchange of intercepts between these branches corresponds to an exchange of stability between the fixed points of $\Phi(t; 1, t_{c})$. Solutions to Eq. (13) provide insights into the diverging mean and variance of $\tau'_{p}$ at the critical point, as the hitting rate can be derived from $\Phi(t; 1, t_{c})$ according to the formula $h(t; 1, x_{c})=-\partial_{t}\Phi(t; 1, t_{c})$ \footnotemark[\value{footnote}]. When $\mu \neq 0$, the solution is a logistic function, so that the hitting rate is exponentially bounded. However, when $\mu=0$, the solution is $\sim(k_{1}^{+}m_{0}t)^{-1}$, meaning that the hitting time distribution becomes heavy-tailed and its moments diverge. This is an example of critical slowing, and is caused by the disappearance of the linear term in $\partial_{t}\Phi(t; 1, t_{c})$. In equilibrium systems, a relationship between critical slowing and diverging susceptibility can be established by considering the Landau potential $V(\Phi)$, which satisfies the law $\partial_{t}\Phi=-\partial_{\Phi}V(\Phi)$. Both the relaxation time following a small variation, and the susceptibility of $\Phi$ to a linearly coupled external field, are inversely proportional to $\partial_{\Phi}^{2}V(\Phi)$. At the critical point, disappearance of the linear term in $\partial_{t}\Phi$ means that $\partial_{\Phi}^{2}V(\Phi)=0$, so that both the relaxation time and susceptibility diverge. In our system, the connection between critical slowing and the susceptibility is more direct, as the mean propagation time defines a lower bound for the susceptibility. Nonetheless, the principle is similar, in that the exchange of stability by the fixed points causes disappearance of the linear term in Eq. (13), allowing the susceptibility to another parameter to diverge.

It is possible to obtain an effective potential of the form $V(\Phi)=-\mu\Phi^{2}/2+k_{1}^{+}m_{0}\Phi^{3}/3$ by integrating Eq. (13) with respect to $\Phi$. However, this result must be interpreted with caution. Thermodynamic equilibrium relies on microscopic reversibility, which does not exist for our parameter $\Phi(t; 1, t_{c})$. Due to the irreversible nature of growth mode switching, $\Phi(t; 1, t_{c})$ can only decrease; thus, the regions of Eq. (13) for which $\partial_{t}\Phi(t; 1, t_{c})>0$ are non-physical, and correspond to the parts of $\hat{h}(s; 1)$ situated at $s<0$ (Fig.~\ref{fig05}(b)). Instead, we propose that the shared feature between our critical point and those observed in equilibrium systems is the presence of a bifurcation. While our bifurcation is the result of irreversible dynamics, bifurcations observed in equilibrium systems are caused by large numbers of microscopically reversible processes. Nonetheless, because these processes result in similar dynamical laws, our model has many features in common with equilibrium phase transitions.

\section{Effects of criticality on the macroscopic self-assembly kinetics}

Autocatalytic biopolymer systems with high $\zeta$ will exhibit a nonlinear response to $\mu$ near the critical point. In this section, we evaluate the effects of this criticality on the macroscopic self-assembly kinetics. We begin by estimating likely $\zeta$ values based on typical biological free energy scales. We find that stabilization of conformational transition states by even modest free energy differences ($<10 k_{B}T$) can result in highly autocatalytic conformational conversion. Thus, different biopolymers are expected to exhibit the full spectrum of autocatalytic behaviors predicted by our model. In systems with high $\zeta$, the critical point has a profound effect on the macroscopic kinetics, inducing rapid changes in the size and abundance of steady-state populations of intermediates, and the concentration-dependent scaling behavior of the formation of mature polymers. These effects are not predicted by existing local equilibrium models \cite{Lomakin1996,Meisl2014,Habchi2018}, and indicate that nonequilibrium criticality may explain widespread experimental phenomena \cite{Gillam2013, Bernacki2009} which one-step nucleation models have not been able to address.

\subsection{Estimation of $\zeta$ from Kramers theory}

Critical behavior arises in the autocatalytic ($\zeta\to1$) limit (Fig.~\ref{fig03}(b)). Although $k_{c}$ is non-zero and $k_{p}$ is finite, meaning that a value of $\zeta=1$ is not strictly possible, $\zeta$ can be asymptotically close to this limit. Realistic $\zeta$ values can be estimated from reaction rate theory. Let us consider the conversion rate of a single type-1 monomer in the AEM domain, situated $n$ monomers from the active end. We represent the conformation of the monomer with a single continuous degree of freedom, $q_{n}$; however, we note that our analysis can be generalized to cases where monomers have additional slow degrees of freedom without changing the conclusions. The free energy $G(q_{n}; q_{n-1}, q_{n+1})$ of a polymerized monomer depends on its own conformation $q_{n}$, and a term accounting for interactions with the adjacent monomers $q_{n-1}$ and $q_{n+1}$. The monomer at position $n$ has free energy minima corresponding to the type-1 and type-2 states, which are situated at $q_{n}=p_{1}$ and $q_{n}=p_{2}$. These minima are separated by a barrier at $q_{n}=p^{\ddagger}$, which corresponds to a conformational transition state. The fact that conversion has a negligible reverse process entails that $G(p_{1}; q_{n-1}, q_{n+1}) \gg G(p_{2}; q_{n-1}, q_{n+1})$, and the effectively two-state nature of our model entails that $G''(p_{1}; q_{n-1}, q_{n+1})$ and $G''(p_{2}; q_{n-1}, q_{n+1})$ are both large. Therefore, the conformation $q_{n'}$ of all non-converting monomers $n' \neq n$ is assumed to be close to $p_{1}$ or $p_{2}$. Monomers situated closer to the active end ($n' < n$) are part of the AEM domain, and must be in the type-1 state ($q_{n'} \approx p_{1}$); monomers situated further from the active end ($n' > n$) may be in either the type-1 or type-2 state ($q_{n'} \approx p_{1}$ or $q_{n'} \approx p_{2}$). Although it is possible to write a Langer equation \cite{Langer1969,Hanggi1990} for the conversion rate which accounts for continuous conformational variation in the monomers at positions $n' \neq n$, the results are hard to interpret. The main reason for this is that such a model predicts dynamics based on the $j$-dimensional free energy function $\sum_{n}G(q_{n}; q_{n-1}, q_{j+1})$, which has $2^{j-1}$ relevant saddle points. Instead, we simplify the analysis by neglecting variation in $q_{n'}$ about the points $p_{1}$ or $p_{2}$. We write the free energy of the monomer at position $n$ as a pair of functions corresponding to the two possible conformations of the monomer at position $(n+1)$:
\begin{subequations}
\begin{gather}
    G_{1}(q_{n}) = G(q_{n}; p_{1}, p_{1}),\\
    G_{2}(q_{n}) = G(q_{n}; p_{1}, p_{2}).
\end{gather}
\end{subequations}
The monomer conformation exhibits ovderdamped Langevin dynamics, so that the corresponding conformational probability density functions $\rho_{1}(q_{n}, t)$ and $\rho_{2}(q_{n}, t)$ satisfy the following Fokker-Planck equation:
\begin{equation}
    \frac{\partial \rho_{i}}{\partial t} = \frac{\partial}{\partial q_{n}}D_{i}\bigg[\frac{\partial \rho_{i}}{\partial q_{n}} + \beta\rho_{i}\frac{dG_{i}}{dq_{n}}\bigg]
\end{equation}
where $i=1,2$ is the conformation of the monomer at position $(n+1)$, $\beta=1/k_{B}T$, and $D_{i}(q_{n})$ is the conformational diffusion coefficient. The conversion rate can be estimated from $G_{i}(q_{n})$ and $D_{i}(q_{n})$ using Kramers theory \cite{Kramers1940,Hanggi1990}. The non-autocatalytic conversion rate is given by the passage rate at the barrier $(q_{n}, q_{n+1})=(p^{\ddagger}, p_{1})$:
\begin{equation}
    k_{c} = \frac{\beta D_{1}(p^{\ddagger})}{2\pi}\big[G''_{1}(p_{1})|G''_{1}(p^{\ddagger})|\big]^{1/2}e^{-\beta\Delta G_{1}^{\ddagger}},
\end{equation}
where $\Delta G_{1}^{\ddagger} = G_{1}(p^{\ddagger})-G_{1}(p_{1})$. Similarly, the autocatalytic conversion rate is given by the passage rate at the saddle point $(q_{n}, q_{n+1})=(p^{\ddagger}, p_{2})$:
\begin{equation}
    k_{c} + k_{p} = \frac{\beta D_{2}(p^{\ddagger})}{2\pi}\big[G''_{2}(p_{1})|G''_{2}(p^{\ddagger})|\big]^{1/2}e^{-\beta\Delta G_{2}^{\ddagger}},
\end{equation}
where $\Delta G_{2}^{\ddagger} = G_{2}(p^{\ddagger})-G_{2}(p_{1})$. Therefore, the definition $\zeta=k_{p}/(k_{c}+k_{p})$ gives the result:
\begin{equation}
    \zeta = 1 - \frac{D_{1}(p^{\ddagger})\big[G''_{1}(p_{1})|G''_{1}(p^{\ddagger})|\big]^{1/2}}{D_{2}(p^{\ddagger})\big[G''_{2}(p_{1})|G''_{2}(p^{\ddagger})|\big]^{1/2}}e^{-\beta\Delta\Delta G^{\ddagger}},
\end{equation}
where $\Delta\Delta G^{\ddagger}=\Delta G_{1}^{\ddagger}-\Delta G_{2}^{\ddagger}$ is the reduction in the free energy barrier of conversion due to autocatalytic effects. It is not immediately obvious whether the pre-exponential ratio would have a value greater or less than $1$. While ordering of the $(n+1)$ monomer might be expected to reduce conformational diffusion, the simultaneous loss of complementarity between the monomers' structures might exert the opposite effect. The effects on $G''(p_{1})$ and $G''(p^{\ddagger})$ are also hard to gauge, and may be biopolymer-specific. In the absence of further information, we set the value of the pre-exponential ratio to $1$, so that $\zeta=1-\exp{(-\beta\Delta\Delta G^{\ddagger})}$, and we examine the relationship between $\zeta$ and $\Delta\Delta G^{\ddagger}$. Eq. (18) predicts that even high $\zeta$ values will be associated with relatively modest free energy differences; for example, $\zeta=0.999 \Leftrightarrow k_{p}\approx10^{3}k_{c}$ implies $\Delta\Delta G^{\ddagger}=6.9\,k_{B}T$, which is well within the range of typical biological free energy scales at $T = 310\,K$. For comparison, formation of a single hydrogen bond in protein has a free energy change of $\sim2\,k_{B}T$ \cite{Ben-Tal1997,Sheu2003}, and estimated free energy barriers for protein folding are mostly in the $2-20\,k_{B}T$ range \cite{Naganathan2005}, accounting for the effects of large numbers of competing, non-additive interactions. In addition, a recent experimental study of biopolymer nucleation by the A\textbeta\space peptide indicated that autocatalytic interactions between polymers reduce the nucleation free energy barrier by $\Delta\Delta G^{\ddagger}=19\,k_{B}T$ \cite{Cohen2018}. Although this form of autocatalysis is different from the form that we consider here, as it depends on interactions between polymers rather than within a polymer, the two processes may share a common physical basis. It is interesting to note that an equivalent $\Delta\Delta G^{\ddagger}$ value in our model would give $\zeta\approx1-10^{-8} \Leftrightarrow k_{p}\approx10^{8}k_{c}$, resulting in highly autocatalytic behavior. Thus, we believe that values of $\zeta$ close to the $\zeta=1$ limit are highly plausible, meaning that biopolymer systems will frequently exhibit high levels of autocatalysis and resulting critical behavior.

\subsection{Nonlinear response to perturbations in the autocatalytic limit}

We now investigate the response of the macroscopic self-assembly kinetics to variations in $\mu/D$, when $\zeta$ is high. To do so, we calculate the time-evolution of the principal moments of the polymer length distribution. Let $f_{i,j}(t)$ represent the concentration of polymers with growth mode $i=1,2$ and length $j$ at time $t$. In parallel with the notation used by \cite{Knowles2009,Cohen2011a}, we define the following principal moments:
\begin{equation}
    P_{i}(t) = \sum_{j=j_{0}}^{\infty}f_{i,j}(t),
\end{equation}
\begin{equation}
    M_{i}(t) = \sum_{j=j_{0}}^{\infty}jf_{i,j}(t).
\end{equation}
The zeroth-order moment $P_{i}(t)$ is the total concentration of polymers with growth mode $i$, and acts as an effective normalization constant for $f_{i,j}(t)$. The first-order moment $M_{i}(t)$ is the effective concentration of monomers incorporated into polymers with growth mode $i$, or the `polymer mass-concentration'. The mean polymer length $L_{i}(t)$ can be obtained by the normalization:
\begin{equation}
    L_{i}(t) = \frac{P_{i}(t)}{M_{i}(t)}.
\end{equation}
Let us begin by examining the effect of the critical point on $M_{1}(t)$ and $L_{1}(t)$. The length distribution $f_{1,j}(t)$ of the population of type-1 species contains contributions from individual polymers initiated at all $t_{0}<t$. If nucleated polymerization begins when $t=0$, then
\begin{equation}
    f_{1,j}(t) = v_{n}\sum_{x=1}^{j}\int_{0}^{t}p_{j,x}(t; t_{0})dt_{0}
\end{equation}
where $v_{n}=k_{n}m_{0}^{j_{0}}$ is the nucleation rate. Thus,
\begin{equation}
    P_{1}(t) = v_{n}\sum_{j=j_{0}}^{\infty}\sum_{x=1}^{j}\int_{0}^{t}p_{j,x}(t; t_{0})dt_{0},
\end{equation}
\begin{equation}
    M_{1}(t) = v_{n}\sum_{j=j_{0}}^{\infty}j\sum_{x=1}^{j}\int_{0}^{t}p_{j,x}(t; t_{0})dt_{0}.
\end{equation}
In Sec. III. A., we obtained numerical solutions for $p_{j,x}(t; t_{0})$ using an rfKMC algorithm. We will now use these solutions to predict $M_{1}(t)$ and $L_{1}(t)$, by applying Eq. (23-24). As we are specifically interested in the response of the system to varying $\mu/D$ at high $\zeta$, we will focus on cases where $\mu/D$ varies through the range $[-2, 2)$, while $\zeta$ is fixed at the value $\zeta=0.999 \Leftrightarrow k_{p}\approx10^{3}k_{c}$. This value is associated with a comparatively small $\Delta\Delta G^{\ddagger}$, and so is physically very plausible (Sec. IV. A.).

\begin{figure*}[t]
\includegraphics[width=17.7cm]{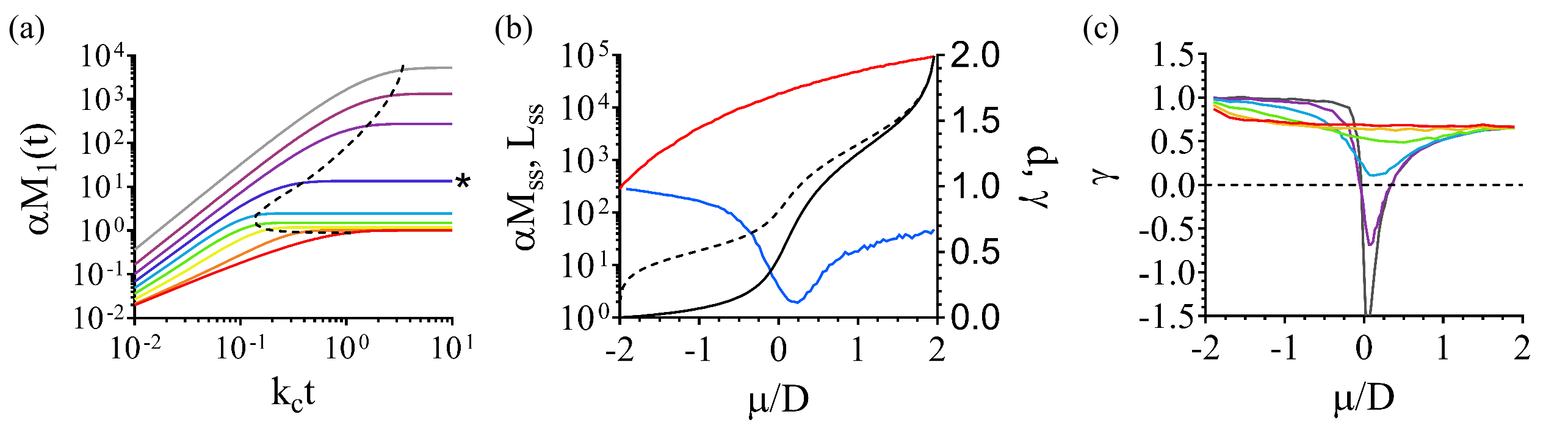}
\caption{\label{fig06} Varying $\mu$ close to the critical point ($k_{p}=10^{3}k_{c}$) results in a highly nonlinear response of the macroscopic kinetics. (a) Time-evolution of the mass of species with a type-1 growth mode, $\alpha M_{1}(t)$, for varying $\mu/D$. Color scale: red, $\mu/D=-2$; orange, $-1.9$; yellow, $-1.5$; green, $-1$; cyan, $-0.5$; blue, $0$ (starred, critical); indigo, $0.5$; purple, $1$; gray, $1.5$. The dashed line plots the $(k_{c}t, \alpha M_{1}(t))$ at which $\alpha M_{1}(t)=0.9\alpha M_{ss}$, for varying $\mu/D$. (b, left axis) Effect of $\mu/D$ on the steady-state mass ($\alpha M_{ss}$; solid black) and length ($L_{ss}$; dashed black) of species with a type-1 growth mode. (b, right axis) Effect of $\mu/D$ on the scaling exponents $d$ (red) and $\gamma$ (blue). (c) Dependence of $\gamma$ on $\mu/D$, for varying $\zeta$. Color scale: red, $\zeta=0$; amber, $\zeta=0.9$; green, $\zeta=0.99$; blue, $\zeta=0.999$; purple, $\zeta=0.9999$; dark gray, $\zeta=0.99999$.}
\end{figure*}

In Fig.~\ref{fig06}, we present the effect of $\mu/D$ on the time-evolution of $\alpha M_{1}(t)$ and $L_{1}(t)$, where $\alpha=k_{c}/v_{n}$ is a normalization constant facilitating the use of relative time $k_{c}t$. The integral $\int_{0}^{\infty}p_{j,x}(t; t_{0})$ converges, allowing both $\alpha M_{1}(t)$ and $L_{1}(t)$ to approach steady-state values $\alpha M_{ss}=\alpha\lim_{t\to\infty}M_{1}(t)$ (Fig.~\ref{fig06}(a) \& (b)) and $L_{ss}=\lim_{t\to\infty}L_{1}(t)$ (Fig.~\ref{fig06}(b)) in the $t\to\infty$ limit. These nonequilibrium steady states reflect the concentrations and size of intermediates likely to be observed in physiological contexts, if the free monomer concentration and chemical environment are stable on timescales $\gg \tau_{\textrm{steady}}$, where $\tau_{\textrm{steady}}$ is the approximate timescale taken to reach the steady state. In experimental contexts, the pre-steady-state behavior is likely to dominate in the self-assembly lag phase if $\tau_{\textrm{inf}} \ll \tau_{\textrm{steady}}$, and the steady-state behavior is likely to dominate if $\tau_{\textrm{inf}} \gg \tau_{\textrm{steady}}$.

We will evaluate the effects of the critical point on both the pre-steady-state kinetics and the steady-state kinetics. The pre-steady-state kinetics appear linear when plotted on double-logarithmic axes (Fig.~\ref{fig06}(a)), indicating that they are described by a power law of the form $\alpha M_{1}(t) \propto t^{d}$. Power laws are often observed in biopolymer self-assembly kinetics where $t$ is small, and $d$ can be interpreted as the number of sequential processes contributing to biopolymer mass accumulation; thus, nucleation without polymerization gives $d=1$, and nucleated polymerization gives $d=2$. To assess whether the pre-steady-state kinetics are affected by the critical point, we calculated $d=\partial\ln{M_{1}(t)}/\partial\ln{t}|_{k_{c}t=0.01}$ for varying $\mu/D$ (Fig.~\ref{fig06}(b)). Although $d$ increases with $\mu/D$, reflecting the increasing importance of polymerization in mass accumulation, there is no clear response to the critical point. This is unsurprising, given that the critical point results from changes in growth mode switching, which is negligible in the pre-steady-state.

In contrast, the steady-state kinetics are strongly affected by the critical point. We begin by examining the effect of the critical point on $\tau_{\textrm{steady}}$. For the purpose of this analysis, we define $\tau_{\textrm{steady}}$ such that $\alpha M_{1}(\tau_{\textrm{steady}})=0.9\alpha M_{ss}$. As shown in Fig.~\ref{fig06}(a), $\tau_{\textrm{steady}}$ initially decreases with increasing $\mu/D$, as both elongation and propagation cooperate to accelerate growth mode switching. However, this trend reverses close to the critical point, causing a delay in the approach to the steady-state and allowing greater accumulation of polymer mass. This delay corresponds to the emergence of a population of large, conformationally mixed intermediates. As a result, both $\alpha M_{ss}$ and $L_{ss}$ exhibit a highly nonlinear response to $\mu/D$ around the critical point (Fig.~\ref{fig06}(a) \& (b)). In physiological and experimental contexts, this means that small variations in the free monomer concentration or solvent composition have the potential to induce large changes in the populations of biopolymer nucleation intermediates. In addition, the total rate of growth mode switching is strongly dependent on the size and nature of the intermediate populations, so the increase in $\tau_{\textrm{steady}}$ will strongly affect the accumulation of mature polymers with a type-2 growth mode.

\subsection{Low concentration-dependence is a dynamical signature of a nonequilibrium critical point}

As discussed in Sec. II. B., a common experimental descriptor of the macroscopic self-assembly kinetics is the characteristic time $\tau_{\textrm{char}}$ at which $M_{2}(\tau_{\textrm{char}})=Cm_{0}$, where $C$ is an arbitrary constant (Fig.~\ref{fig02}(b)). The concentration-dependence of $\tau_{\textrm{char}}$ can be described using a scaling exponent $\gamma$, which has the following definition:
\begin{equation}
    \gamma = -\frac{\partial\ln{\tau_{\textrm{char}}}}{\partial\ln{m_{0}}}.
\end{equation}
In cases where $C$ is sufficiently small ($0 < C \lessapprox 0.5$), mechanistic conclusions can be drawn from the value of $\gamma$ \cite{Ferrone1999}. For example, the Oosawa model predicts $\gamma=n_{c}/2$ \cite{Oosawa1975}, where $n_{c}$ is the nucleation order, and the nucleation-conversion-polymerization (NCP) model, which assumes instantaneous growth mode switching, predicts $\gamma=n_{c}/3$ \cite{Garcia2014}. To illustrate the impact of the abrupt change in $\tau_{\textrm{steady}}$ on the macroscopic self-assembly kinetics, we predict $\gamma$ values based on our simulations. We begin by observing that:
\begin{equation}
\begin{split}
    \frac{df_{2,j}(t)}{dt} =&\;v_{n}(k_{1}^{-}+k_{p})\int_{0}^{t}p_{j,1}(t; t_{0})dt_{0}\\
    +&\;k_{2}^{+}m_{0}[f_{2,j-1}(t)-f_{2,j}(t)]\\
    -&\;k_{2}^{-}[f_{2,j+1}(t)-f_{2,j}(t)].
\end{split}
\end{equation}
Our rfKMC simulations were carried out in the $k_{1}^{-}\to0$ limit, meaning that the following conservation principle applies:
\begin{equation}
    P_{2}(t) = v_{n}t - P_{1}(t).
\end{equation}
To simplify our analysis, let us consider the case where $k_{2}^{+}m_{0} \gg M_{ss}/\tau_{s}$, so that the predominant contribution to the type-2 polymer mass comes from elongation, rather than conversion of intermediates. Therefore,
\begin{equation}
    M_{2}(t) \approx k_{2}^{+}m_{0}\int_{0}^{t}P_{2}(\tau)d\tau.
\end{equation}
In the pre-steady-state ($t \ll \tau_{\textrm{steady}}$), $P_{1}(t)$ is approximately linear, so both $P_{2}(t)$ and $M_{2}(t)$ will exhibit higher-order scaling with time. In the steady-state ($t \gg \tau_{\textrm{steady}}$), $P_{1}(t)$ is approximately constant, meaning that $P_{2}(t)$ and $M_{2}(t)$ will exhibit linear and quadratic scaling with time, respectively. It is more convenient to express concentrations in relative units than it is to choose arbitrary values of the rate constants. Therefore, we calculate $\alpha'M_{2}(t)$ using Eq. (23, 27-28), where $\alpha'=(k_{1}^{+}/k_{p})^{j_{0}+1}(k_{c}^{2}/k_{n}k_{2}^{+})$ is a normalization constant, and set $C=0.01/\alpha'$. From the definition of $\mu$ and $D$, we find that
\begin{equation}
    \frac{\partial\ln{\big(\frac{\mu}{D}\big)}}{\partial\ln{m_{0}}} = \frac{4-\big(\frac{\mu}{D}\big)^{2}}{4\big(\frac{\mu}{D}\big)}.
\end{equation}
Thus, under circumstances where the rate constants are kept fixed, variation in $\mu/D$ can be attributed to changes in the free monomer concentration $m_{0}$. This means that we are able to determine $\gamma$ by calculating $\tau_{\textrm{char}}$ for varying, finely spaced $\mu/D$ values, estimating the logarithmic derivative with respect to $\mu/D$ from the corresponding finite difference equation, and then multiplying by the right-hand side of Eq. (29). In Fig.~\ref{fig06}(c), we present the results obtained for various $\zeta$. At low $\zeta$, $\gamma$ monotonically transitions between the limits $\lim_{\mu\to-2}\gamma=1$ and $\lim_{\mu\to2}\gamma=2/3$, which agree with the predictions of \cite{Oosawa1975} and \cite{Garcia2014} discussed above. However, a minimum in $\gamma$ appears at high $\zeta$, which starts at the point $\mu/D=2$ and moves towards $\mu/D=0$ as $\zeta\to1$. This depression occurs because $\mu/D$ scales with $m_{0}$, so that increasing $m_{0}$ causes a reduction in growth mode switching, inhibiting further acceleration of type-2 mass accumulation. In highly autocatalytic systems ($k_{p} > 10^{3}k_{c}$), the effect is severe enough to cause a temporary reversal in concentration-dependence. As the minimum $\gamma$ decreases, the drop in $\gamma$ becomes increasingly sharp about the point $\mu/D=0$. Both the low concentration-dependence ($\gamma<0$) and sharp drop in $\gamma$ predicted by our model are commonly observed in experimental studies, but cannot be explained by existing models \cite{Lomakin1996,Meisl2014,Habchi2018}. Furthermore, while existing models invoke local equilibria to explain low $\gamma$ \cite{Lomakin1996,Meisl2014,Habchi2018}, our model predicts a low $\gamma$ purely as a result of the dynamics. We thus propose that a low concentration-dependence in biopolymer self-assembly may represent the dynamical signature of nonequilibrium critical point.

\section{Discussion}

Biopolymer self-assembly is a fundamental mechanism of biological organization, and plays a key role in disease. Experimental and computational studies have revealed that monomers often populate metastable conformational states during the process of self-assembling to form a biopolymer \cite{Uversky2010, Straub2011, Colletier2011, Kodali2001}. As a result, nucleation of a biopolymer often proceeds via one or more metastable intermediates, whose constituent monomers are conformationally distinct from those within the mature polymer. In addition, monomers within the same intermediate often exhibit considerable conformational heterogeneity. This complexity has challenged our ability to extract detailed mechanistic information at the molecular level. Current mathematical models either neglect the conformational heterogeneity of the assembled monomers entirely, or assume that monomers within the same intermediate rapidly assume the same conformational state, so that intermediates with a mixed composition are not observed. However, mixed intermediates are commonly observed in experiments and simulations, indicating that mathematical models which neglect these dynamics are only applicable under highly simplified experimental conditions \cite{Uversky2010}. Moreover, this conformational diversity is responsible for many of the most important biological behaviors of biopolymers \cite{Uversky2013, Babu2011, Vavouri2009}. One notable example is the role of intermediates in the assembly of amyloid fibres from protein monomers. The conformational heterogeneity of amyloid self-assembly intermediates allows them to interact with a diverse range of other biomolecules \cite{Miller2010}. This disrupts cellular function and is causative in the majority of degenerative conditions of aging, including Alzheimer's and Parkinson's diseases \cite{Lambert1998, Lesne2006, Benilova2012, Ono2009, Xue2010, Shin2017}.

In this paper, we address this issue by introducing a minimally complex process by which the self-assembled monomers are able to exhibit conformational heterogeneity, and change conformation over time. This simple addition causes our model to exhibit rich, critical dynamics. As a result, the macroscopic self-assembly kinetics display features such as a highly nonlinear dependence of steady-state intermediate populations on the self-assembly conditions, and loss or even inversion of the concentration-dependence of the self-assembly rate of the stable phase. These phenomena are commonly observed in experimental studies \cite{Gillam2013, Bernacki2009}, but are not explained by existing models. Thus, for the first time, the initiation-propagation (IP) model proposed here allows quantitative predictions to be made regarding the concentration of conformationally heterogeneous intermediates.

\subsection{Competing dynamics in morphologically diverse intermediates}

While polymeric intermediates are commonly observed in biopolymer self-assembly, a diverse range of other intermediate morphologies are also observed, including small prenucleation clusters with an apparently random organization of constituent monomers \cite{Ahmed2010,Bitan2003}, large spheroidal intermediates \cite{Lomakin1996,Sabate2005,Yong2002}, and species possessing fractal geometry which is evident at a macroscopic scale \cite{Murr2005}. In addition to applying to polymeric intermediates, we expect the conclusions of our study to generalize to these other diverse geometries. To make this generalization, we note that propagation involves incorporation of monomers into a growing stable phase by autocatalytic conversion of condensed monomers in the intermediate phase. Thus, the dimensionality of propagation is defined by that of the stable phase. In an $n$-dimensional intermediate, propagation will manifest as linear growth of the stable phase through the intermediate, with the growth mode switching when an autocatalytic end reaches the solvent. This form of propagation is well-supported by experimental observations of spheroidal intermediates \cite{Zhu2002,Galkin2007,Luo2014}, which are often large enough to be observed by light microscopy techniques. In these studies, the stable polymeric phase appears to originate within the intermediate by conformational rearrangement of a subset of the monomers, and then emerge from the surface of the intermediate. When the active end reaches the solvent, the dramatic change in the physicochemical environment, and the availability and conformational state of nearby monomers, would be expected to induce a marked change in the growth behavior of the polymeric phase analogous to the switch from propagation ($k_{c}+k_{p}$) to type-2 growth ($k_{2}^{+}m_{0}$) behavior in our polymer model.

The fact that polymerization is likely to manifest as linear growth-like behavior in multidimensional intermediates means that we expect IP models involving such intermediates to reduce to pseudo-1D absorbing boundary problems. A transition from a regime where $P_{e}=0$ to one where $P_{e}>0$ is a general feature of such problems \cite{Kendall1948}, so we expect similar transitions for diverse intermediate geometries. It is also important to note that our critical point is conserved in the continuous limit of $x$, ie. the case where the intermediate is so large as to behave like a continuous phase, rather than a discrete collection of monomers. In this limit, we obtain the classical results for first-passage of Brownian motion with drift \cite{Cox1965}, and a jump in $\partial_{\mu}P_{e}(x_{c})$ occurs about the point $\mu=0$ \footnotemark[\value{footnote}]. Besides the discrete polymer example considered in this paper, IP mechanisms with a number of other intermediate geometries are likely to be mathematically tractable. For example, in the case of spherical intermediates with a continuous-like composition, the problem considered in Sec. II. can be recast as a continuous advection-diffusion (Eq. (4)) along a 1D trajectory towards a spherical absorbing boundary. In the simplified case of radial propagation, we have performed a quick calculation to estimate the rate at which large spherical intermediates give rise to a growing stable polymers \footnotemark[\value{footnote}]. While this rate is proportional to the volume of the intermediates in the cooperative regime, it is proportional to their surface area in the competitive regime. This behavior closely resembles the way in which initiation sites for successful propagation events are restricted to the ends of polymeric intermediates in the competitive switching regime. Thus, a more general principle may exist, in which the competition between growth and propagation dynamics creates two distinct regimes. In one, a polymerizing stable phase may originate from anywhere within the volume of the intermediate; in the other, the sites where such a phase may originate are restricted to the boundary of the intermediate. The fact that a critical point separating these regimes appears to be a widespread feature of two-step biopolymer nucleation pathways underlines the more general principle that the prerequisites for critical behavior are relaxed in far-from-equilibrium systems.

In addition to being compatible with a wide range of intermediate morphologies, our IP mechanism also \textit{predicts} significant morphological diversity. The incorporation of a single alternative monomer conformation means that polymers of a given length may exhibit a large number of different sequences of type-1 and type-2 monomers. Because the conformation of a monomer affects its mechanical properties, we would expect these different sequences to result in diverse morphologies. In this study, we have primarily focused on the mechanism of growth mode switching, and have ignored aspects of the conformational conversion dynamics which are not relevant to this problem. However, in future studies it may be possible to rigorously predict changing polymer morphologies by obtaining analytical solutions to describe the progressive conversion of monomers throughout the polymer. As an example of how polymer composition might affect morphology, we observe that monomers in the type-1 and type-2 states are likely to exhibit differing degrees of conformational ordering. It is most often assumed that the intermediate is less ordered than the stable species, so that conformational conversion can be regarded as an ordering transition. In this case, the conformational degeneracy of the type-1 state would be expected to result in greater flexibility for type-1 domains. As the monomers within a polymer converted to the type-2 state, we would expect to see a progressive increase in the persistence length $l_{p}$. However, in cases where the persistence length of the type-1 and type-2 monomers are markedly different, presence of a small number of type-1 monomers would be expected to strongly affect the overall morphology of the polymer. Thus, polymers with a small but significant type-1 content might exhibit radically different morphologies from mature polymers, as is often observed experimentally \cite{Kodali2001,Meinhardt2007,Pedersen2010}.

It would also be possible to introduce further conformational diversity to our model, by introducing additional metastable conformations for the polymerized monomers. While analysis of such a model would not be trivial, additional conformational freedom and the presence of other slow dynamics might be expected to yield even richer behaviors than those presented here. However, it is useful to note that structural data from solid-state nuclear magnetic resonance (NMR) studies suggest that monomers within amyloid fibrils occupy a limited range of stable or metastable conformations \cite{Tycko2011,Iadanza2018}, while high-resolution microscopy data have revealed a diverse range of morphologies \cite{Meinhardt2007,Jimenez2002}. Therefore, even in amyloid systems where the free monomer is often highly conformationally degenerate, the range of accessible conformations seems to remain limited within the polymer. Thus, a simple model in which polymerized monomers occupy one of two conformations may be sufficient to account for much of the observed morphological diversity.

\subsection{Explicit conversion dynamics are essential to a general two-step biopolymer nucleation model}

The inclusion of explicit conformational conversion dynamics also allows our IP model to exhibit a generality not seen in other two-step nucleation models. Propagation of the stable phase is not assumed to be instantaneous, and can occur on timescales similar to either polymerization of the intermediate or non-autocatalytic conversion (Fig.~\ref{fig02}). As a result, the dynamical behavior of our model depends on two key parameters, $\mu/D$ and $\zeta$. The first of these, $\mu/D$, describes the competing effects of propagation and polymerization on the time-dependence of growth mode switching. When $\mu/D<0$, the stable phase propagates through the intermediate faster than the intermediate phase can polymerize, causing accelerated switching. When $\mu/D>0$, polymerization is rapid and reduces the likelihood of successful propagation. The second parameter, $\zeta$, compares the timescale for propagation due to autocatalytic conversion with the timescale for non-autocatalytic conversion. When $\zeta=0$, nearest-neighbor interactions between monomers do not promote conformational conversion, so that the rate of growth mode switching is independent of the length of the polymer. When $\zeta=1$, nearest-neighbor interactions strongly promote conversion, so that the timescale for propagation is much faster than that of non-autocatalytic conversion, and the rate of growth mode switching is accelerated.

Existing models which neglect the conversion dynamics are not able to consider the variation accounted for by $\mu/D$ and $\zeta$, and thus make various assumptions regarding the rate at which the new growth mode emerges. While certain models \cite{Auer2012,Saric2016} have correctly supposed a connection between intermediate size and the rate of emergence of a growing stable phase (eg. $k_{c}\langle\tau_{s}\rangle\approx1/j$), the lack of competing dynamics means they are restricted to a limited range of points close to the top left ($\zeta\to1$, $\mu/D<0$) of our phase diagram (Fig.~\ref{fig03}(a)). In contrast, other models have assumed that there is no relationship between intermediate size and the switching rate \cite{Vitalis2011,Garcia2014} (ie. $k_{c}\langle\tau_{s}\rangle\propto1$). Justifications for this behavior include irrelevance of size due to the lack of a growth process, or a scenario in which monomer conversion is highly cooperative. While the former is plausible in specific cases where growth of intermediates is slow, the latter is unlikely to be widely observed, as intermediates are typically either low-dimensional or have highly degenerate monomer interactions. However, a sufficient level of cooperativity may occur in exceptional cases where monomers within intermediates exhibit a limited variety of interactions with a large number of neighbors.  In our model, length-independence arises in two scenarios: when conversion is entirely non-autocatalytic so that adjacent monomers convert independently ($\zeta=0$); and when growth is sufficiently rapid that growth mode switching is inhibited ($\mu/D\to2$). Thus, the behavior assumed by \cite{Vitalis2011,Garcia2014} will be observed at points on our phase diagram corresponding to these limits (Fig.~\ref{fig03}(a)). It is important to emphasize that we expect length-independence to be a somewhat unusual phenomenon. Even stabilization of the conformational transition state by a small free energy ($\sim1\;k_{B}T$) would result in a significant $\zeta$ value, and in the vast majority of our phase space some level of reduction in $k_{c}\langle\tau_{s}\rangle$ is observed (Fig.~\ref{fig03}(a)).

We have seen that existing models of two-step biopolymer nucleation describe a limited range of possible scenarios, which are restricted to points close to three lines ($\zeta\to1$ given $\mu/D<0$; $\zeta=0$; and $\mu/D\to2$) on the limits of our phase diagram (Fig.~\ref{fig03}(a)). As a result, these models exhibit markedly different behaviors which appear difficult to reconcile. By explicitly considering the dynamics by which conversion of monomers within a polymeric intermediate leads to emergence of a new growth mode, we are able to explore a much broader region of parameter space, and show that these apparently contrasting scenarios are in fact different limits of the same underlying reaction scheme. In addition, we find that these previously unexplored regions contain rich phase behavior which is not anticipated based on the behavior observed in the limits. Therefore, our model has the capacity to explain a wide variety of `poorly behaved' experimental or physiological systems which are not described by existing models, such as those with mixed intermediates or low concentration-dependence (Sec. IV. B. \& C.). Moreover, by classifying existing two-step nucleation models in the context of our phase diagram (Fig.~\ref{fig03}(a)), we have shown that these models occupy disparate regions of phase space which can only be united by considering the competitive dynamics accounted for by $\mu/D$ and $\zeta$. Therefore, while our model still does not describe some of the more complex scenarios that have been suggested \cite{Lee2017}, we have shown that explicit conversion dynamics are essential features of a general model of two-step biopolymer nucleation. We thus believe that our study represents a crucial step towards such a model.

\subsection{Biological importance of the nonequilibrium critical point}

In addition to generalizing existing models, the inclusion of an explicit conformational conversion process allows our model to predict a nonequilibrium critical point. While critical points have previously been described in biopolymer systems, the possibility of nonequilibrium criticality has largely been overlooked. Instead, the most commonly suggested critical point is the critical micellar/oligomer concentration (CMC) \cite{Lomakin1996,Sabate2005,Hasecke2018}, which is the concentration at which the free monomer and intermediate are in local equilibrium, such that $m_{0} = k_{1}^{-}/k_{1}^{+}$. CMC models typically assume that formation of intermediates causes rapid depletion of the free monomer, so that $k_{1}^{-}/k_{1}^{+}$ represents an upper limit on the free monomer concentration; as a result, saturation behavior occurs. In this paper, we consider a different scenario. We focus primarily on the case where the free monomer depletes slowly, as occurs in physiological circumstances where there is a monomer source, or in many experimental situations \cite{Varela2018,Narayan2012,Narayan2014,Cremades2012}. Thus, while our basic model (Fig.~\ref{fig01}) has the required processes to produce a CMC-type effect, we are primarily interested in the case where the free monomer is supersaturated with respect to the intermediate ($m_{0}>k_{1}^{-}/k_{1}^{-}$), and thus out of local equilibrium. Under these circumstances, a second critical point emerges at $m_{0}=(k_{1}^{-}+k_{p})/k_{1}^{+}$ (Fig.~\ref{fig03} \& \ref{fig05}), at a higher concentration than the CMC. On our nonequilibrium phase diagram (Fig.~\ref{fig03}a), the CMC would manifest as a vertical line (ie. constant $\mu/D$) situated to the left of the nonequilibrium critical point. The difference in concentration between these two points will be given by $k_{p}/k_{1}^{+}$. Although we set $k_{1}^{-}=0$ in our simulations to focus primarily on the nonequilibrium critical point, our mathematical analysis is valid for more general $k_{1}^{-}$ subject to the condition that destabilization of intermediates is negligible. Thus, the CMC and the nonequilibrium critical point described in this paper are compatible so long as the concentration difference between the two is sufficiently large. This situation is highly relevant to nonequilibrium scenarios where the dissociation of monomers from the intermediates is relatively slow, as occurs in Alzheimer's and Parkinson's diseases \cite{Cremades2012,Narayan2012,Narayan2014}. In cases where $k_{p} \ll k_{1}^{-}$, the CMC and nonequilibrium critical point will become close to one another on the phase diagram, and the presence of two non-negligible absorbing boundaries in Eq. (1) will further complicate the dynamics of self-assembly and conversion. A rigorous analysis of this effect is both mathematically involved and outside the scope of this paper, and so has been left as a subject for future work.

It is interesting to note that in cases where physiological and experimental systems exhibit slow variation in free monomer levels over time, the phase behavior of biopolymer self-assembly will vary accordingly. Experimental systems are likely to exhibit a monotonic depletion of the free monomer. In this case, $\mu/D$ will decrease over time, causing the system to pass the nonequilibrium critical point \textit{en route} to the CMC. The formation of stable polymers due to growth mode switching will then cause further monomer depletion, resulting in either destabilization or switching of the remaining intermediates. Thus, accelerated growth mode switching will occur when the free monomer concentration passes the nonequilibrium critical point at $m_{0}=(k_{1}^{-}+k_{p})/k_{1}^{+}$, with destabilization of the remaining intermediates occurring shortly after. This sort of behavior has been observed for the amyloid-\textbeta\;(A\textbeta) peptide. Under conditions of high supersaturation, Chimon \textit{et al.} \cite{Chimon2007} described the formation of large, spheroidal intermediates by A\textbeta, similar to those discussed in Sec. V. A. Over time, these intermediates progressively acquired secondary structure content similar to the stable polymeric phase, but growing, stable polymers only appeared following depletion of the free monomer. This behavior is strongly indicative of a transition from the competitive switching regime to the cooperative regime, as described by our model.

The biological implications of the nonequilibrium critical point extend beyond the biopolymer self-assembly process itself. The rapid change in steady-state behavior when $\mu/D\approx0$ (Fig.~\ref{fig06}) means that small changes in free monomer concentration or the physicochemical environment will have a pronounced effect on the population of intermediates. This provides a way for biological systems to rapidly switch between producing stable polymers and metastable intermediates with distinct biological activity. This sort of switch-like regulation is typical of biological systems \cite{Uversky2008,Babu2011}, and may for example reflect the importance of controlling the population of potentially toxic species in the assembly of amyloids \cite{Otzen2008}. Thus, while previous work is restricted to specific sets of conditions, our model encompasses a much wider context. This makes it possible to predict the response of the system to a wide range of stimuli, including changes in internal conditions during the course of a reaction or external stimuli such as mutations and temperature changes, and to interpret the possible role that particular variations in self-assembly conditions play in healthy functioning and disease.

Biological systems are typically maintained far-from-equilibrium, with free monomer concentrations often highly supersaturated in relation to their respective polymers \cite{Varela2018}. However, existing models of two-step biopolymer nucleation neglect the dynamics which arise under these nonequilibrium conditions. The requirements for criticality are relaxed under nonequilibrium conditions, and many of these neglected dynamics have the potential to produce complex, biologically relevant behaviors, such as we observe in our IP model. As discussed above, these behaviors may explain experimental observations which are not predicted by existing models, and rationalize the role of physiological conditions in human disease. As a result, we propose that the role of nonequilibrium critical phenomena in physiological biopolymer self-assembly has been severely underestimated.

\subsection{Relationship to autocatalytic secondary nucleation}

The model developed here can be applied to any two-step biopolymer nucleation process. A fundamental aspect of our model is its ability to predict mixed intermediates consisting of self-assembled monomers in two distinct conformations, resulting in diverse intermediate morphologies. Conversion of a self-assembled monomer from the metastable type-1 state to the stable type-2 state occurs either non-autocatalytically, with rate $k_{c}$, or autocatalytically with rate $k_{c}+k_{p}$. Thus, the stable phase is able to initiate locally within the intermediate, and then propagate autocatalytically to the active end. This type of intramolecular structural propagation has been extensively documented in the biochemical literature in models of multi-subunit proteins \cite{Papaleo2016,Chimon2007,Luo2014,Auer2008,Kumar2009,Chen2015}, and is responsible for the nonequilibrium criticality that our model exhibits.

A different type of autocatalysis which is often discussed in models of biopolymer nucleation is secondary nucleation, the process by which the surface of an existing stable biopolymer catalyzes the formation of additional biopolymers by heterogeneous nucleation \cite{Ferrone1980,Ruschak2007,Cohen2013,Michaels2018}. Thus, while the proposed IP mechanism can be considered a form of intra-polymer autocatalysis, secondary nucleation is an example of inter-polymer autocatalysis. Secondary nucleation has been used to explain features such as exponential ($M_{2}(t) \propto \exp(\kappa t)$) rather than quadratic ($M_{2}(t) \propto t^{2}$) accumulation of the stable phase, and the formation of significant concentrations of toxic intermediates during the `growth phase' (ie. when $t\approx\tau_{\textrm{inf}}$) \cite{Cohen2013}. Given the prominence of secondary nucleation models in the protein polymerization literature, it is pertinent to question how they are mechanistically related to our IP model. At the microscopic level, the process of secondary nucleation results from the alignment of free monomers to the surface of the mature biopolymer \cite{Tornquist2018}. In the context of our model, this alignment could promote initiation of type-1 polymers which then detach from the mature polymer and independently undergo growth mode switching. This would manifest as an increase in the initiation rate $v_{n}$. There is also the possibility that interactions between conformationally ordered monomers in the stable phase and disordered monomers in the nascent intermediate promote ordering of the latter. This would create a more tightly coupled process in which conversion and propagation are accelerated, resulting in an increase in $k_{c}$ and $k_{p}$. Thus, it is possible that the IP mechanism and secondary nucleation share a common mechanistic basis, and are simply different examples of a common tendency for autocatalytic conformational change induced by interactions between monomers. All these effects are easy to incorporate into the framework of our IP model, and we anticipate that future studies which combine the IP mechanism with inter-polymer autocatalysis may be able to gain deeper insights into the microscopic mechanism of secondary nucleation.

\subsection{Competing dynamics can explain low concentration-dependences}

One of the key experimental validators for biopolymer assembly models is their ability to accurately describe the concentration-dependence of the process and relate this to theory \cite{Michaels2016, Bernacki2009, Carbonell2018, Gillam2013,Lee2017,Serio2000,Oosawa1962,Oosawa1975,Cohen2013,Straub2011}. The Oosawa model \cite{Oosawa1962, Oosawa1975} was first to ascribe a physical meaning to the value of $\gamma$ (Eq. (25)), the scaling exponent for the concentration-dependence of accumulation of the stable polymer phase (Sec. IV. C.). The Oosawa model predicts that $\gamma = n_{c}/2$, where $n_{c}$ is the effective order of nucleation and often represents the minimum size for a polymer to be stable or metastable (ie. $j_{0}$) \cite{Ferrone1999}. While $n_{c}$ is often interpreted as reflecting the number of monomers within the critical nucleus ($n^{\ast}=n_{c}-1$) \cite{Ferrone2015}, which is the least stable species during the self-assembly pathway, this explanation has been challenged by the widespread observation of $\gamma<1$ \cite{Gillam2013}. Such low values of $\gamma$ would imply $n^{\ast}<1$, which is physically implausible \cite{Bernacki2009}. In addition, many biopolymer systems with higher $\gamma$ still exhibit much lower concentration-dependences than would be expected, based on biophysical predictions of their $n^{\ast}$ \cite{Auer2012,Bieler2012,Lee2017}. Some authors have proposed that saturation effects, such as the presence of a saturable catalytic surface which induces heterogeneous nucleation, may explain the low concentration-dependences \cite{Habchi2018, Meisl2014, Saric2016b}. Indeed, $\gamma<1$ is generally observed under conditions where $\gamma$ appears to saturate with increasing monomer concentration. However, under some experimental conditions the concentration-dependence drops further to $\gamma<0.5$ \cite{Gillam2013}. This is not only incompatible with the Oosawa model, but it also occurs under circumstances where secondary nucleation remains active, meaning it is incompatible with saturation models. To our knowledge, the nucleation-conversion-polymerization (NCP) \cite{Garcia2014} model is the only model in the literature which produces low $\gamma$ without invoking a saturation effect. In the NCP model, low $\gamma$ values are explained in the context of a `cascade nucleation system', where sequential conversion of $N-1$ pre-fibrillar intermediates results in a concentration-dependence of $n_{c}/(N+1) \leq \gamma \leq n_{c}/2$. In this model, although the number of intermediate types can be infinite, their conversion occurs as a single-step process without considering the underlying dynamics, and the intermediates do not exhibit a growth process. As a result, the conversion mechanism of an NCP intermediate corresponds to the scenario predicted at the point $(\mu/D, \zeta)=(-2, 0)$ on our nonequilibrium phase diagram (Fig.~\ref{fig03}(a)).

The IP mechanism proposed here generalizes the conversion mechanism considered in the NCP model to explicitly consider the dynamics by which intermediates grow and individual monomers convert. As a result, our model makes predictions regarding the pre-steady and steady-state scaling behaviors of intermediates and mature polymers which can easily be generalized to the case of multiple intermediates in the same manner as the NCP model. However, the inclusion of additional dynamics allows our model to explore a broader region of parameter space, leading to our observation of sharp changes in $\gamma$ near the critical point (Fig.~\ref{fig05}(b) \& (c)). This results in local depression of the concentration-dependence, which is more pronounced and occurs more abruptly at higher $\zeta$. High $\zeta$ values correspond to systems in which autocatalysis plays a dominant role in conformational conversion, and can be achieved on comparatively modest free energy scales (Sec. IV. A.). This reduction in concentration-dependence allows our far-from-equilibrium model to violate constraints on the concentration-dependence of existing models over a broad range of $\mu/D$ values, and this effect is likely to be further enhanced when saturation effects are also active. Under many experimental conditions, a value of $\gamma<1$ is thus expected, and is simply a result of the competing dynamics. In this study, the competing processes are polymerization of the metastable intermediate, and propagation of the stable conformational state through this intermediate. However, models which explore other nonequilibrium dynamics are also likely to produce such an effect. Therefore, our work highlights the potential that far-from-equilibrium self-assembly models with competing dynamics have to resolve outstanding issues relating to low concentration-dependences.

\subsection{Interpretation of the nucleus size}

Fig.~\ref{fig06}(c) shows how $\gamma$ varies with $\mu/D$ for different $\zeta$ values. While both $\mu/D$ and $\zeta$ are likely to depend on the physicochemical environment under which self-assembly occurs, and the underlying properties of the biopolymer system, $\mu/D$ exhibits an additional dependence on the free monomer concentration $m_{0}$, which can easily be varied in experimental contexts. It is useful to note that different $\zeta$ values result in highly characteristic changes in concentration-dependence (Fig~\ref{fig06}(c)), so that the $\zeta$ value intrinsic to a particular biopolymer system can be determined by experimentallly varying $\mu/D$ and measuring the changing concentration-dependence of assembly of the stable polymer. Thus, while it is possible to determine $n_{c}$ from the values of $\gamma$ observed in the limits where $\mu/D\to\pm2$, our model predictions provide an easier alternative. Instead, we anticipate that it will be possible to determine $n_{c}$ and $\zeta$ from the characteristic curve shape with which $\gamma$ varies at intermediate values of $\mu/D$. This may be preferable as it eliminates the need to explore extreme values of $\mu/D$, which is challenging due to lack of sensitivity of experimental methods at low concentrations, and the likely presence of additional competing processes at high concentrations. This approach can also be generalized to cases similar to the NCP model \cite{Garcia2014} in which multiple intermediates are present. In this scenario, we would typically expect the Oosawa behavior ($\gamma=n_{c}/2$) to be recovered in the $\mu/D\to-2$ limit, and the NCP-type behavior ($\gamma=n_{c}/(N+1)$) to be recovered in the $\mu/D\to2$ limit, as we have observed in Sec. IV. C. The inclusion of extra intermediates would also be expected to introduce additional critical points, which could be experimentally observable.

It is also important to note that, while our current model identifies $n_{c}$ as being equal to the minimum size of a metastable intermediate $j_{0}$, future models which consider additional complexities of the conversion dynamics may arrive at different interpretations for $n_{c}$. For example, a previous coarse-grained simulation study \cite{Saric2016} which investigated a scenario similar to our cooperative switching regime identified $n_{c}$ with the minimum size at which conversion can initiate within an intermediate, rather than the absolute minimum size of an intermediate. While we take these sizes to be the same in our model, future work which treats these quantities as different may provide analytical justification for this result.

\section{Conclusions}

We have developed a minimal model of two-step biopolymer nucleation which accounts for conformational conversion of metastable intermediates by an initiation-propagation (IP) mechanism. In our framework, the stable polymer phase initiates locally within the intermediate, and propagates by autocatalytic nearest-neighbor interactions between the assembled monomers. This mechanism is well-supported by experimental studies showing progressive ordering of intermediates on slow timescales \cite{Chimon2007,Auer2008,Li2008,Kumar2009,Luo2014,Chen2015}, and computational studies suggesting an autocatalytic mechanism for structural conversion of the monomers \cite{Ding2005,Lipfert2005,Auer2008,Auer2012}. The inclusion of conformational ordering on timescales comparable to condensation of the intermediates causes our model to exhibit rich dynamics and a nonequilibrium critical point, which are not predicted by models with an implicit ordering step \cite{Vitalis2011,Auer2012,Garcia2014,Saric2016,Lee2017}. Furthermore, by explicitly considering the dynamics of conformational conversion, our framework unifies existing models, and shows that their apparently unconnected behaviors occur as different limits of a single underlying mechanism. The IP mechanism has the potential to explain a wide variety of experimental behaviors which are not predicted by existing models, such as the formation of large, conformationally mixed intermediates \cite{Zhu2002,Chimon2007,Galkin2007,Luo2014}, and nucleation of the stable polymer phase at a rate independent of the free monomer concentration \cite{Gillam2013}. In our model, these phenomena arise naturally as a result of the competing dynamics of self-assembly and conversion, resolving the apparent paradox that concentration-independent nucleation rates suggested critical nuclei with a non-physical size \cite{Bernacki2009}.

Biopolymer nucleation intermediates are responsible for toxicity in a huge range of human diseases \cite{Lambert1998,Lesne2006,Ono2009,Benilova2012,Xue2010,Shin2017}, and the nonequilibrium critical point predicted by our model may explain the extreme sensitivity of many polymerization disorders to small changes in self-assembly conditions \cite{Snell1993,Lemere1996,Wadsworth2004}. As a result, our model provides a deeper understanding of the origin of these toxic intermediates, and suggests promising therapeutic strategies to treat diseases caused by their formation. Moreover, future theoretical studies which build on the results presented here are likely to yield analytical solutions for the size distribution and composition of nucleation intermediates, which will provide a powerful tool in therapeutic development. Biopolymers are also increasingly exploited by materials scientists \cite{Wang2016,Gazit2007,Wei2017}, and the development of a general model of their nucleation will allow us to regulate their formation and prevent unwarranted side reactions. In this paper, we have shown that explicit conformational conversion dynamics, such as those addressed by our initiation-propagation mechanism, are essential to such a theory. We thus believe that our work represents a crucial step towards a general theory of two-step biopolymer nucleation, with important implications for both experiments and human disease.\newline

\begin{acknowledgments}
This work was funded by the BBSRC (grant number BB/P002927/1). AIPT gratefully acknowledges financial support from the BBSRC and the University of Sheffield. BC would like to thank the hospitality of MPIPKS, Dresden where part of this work was completed.
\end{acknowledgments}


%

\end{document}